\documentclass[iop]{emulateapj}

\usepackage{amsmath, natbib, ulem}

\newcommand{\kpch}{{\rm kpc}\, h^{-1}}
\newcommand{\Mpch}{{\rm Mpc}\, h^{-1}}
\newcommand{\fesc}{f_{\rm esc}}
\newcommand{\Luv}{L_{\rm UV}}
\newcommand{\Muv}{M_{\rm UV}}
\newcommand{\Mab}{M_{\rm AB}}
\newcommand{\Mstar}{M_\star}
\newcommand{\Msun}{M_\odot}
\newcommand{\Msunh}{M_\odot h^{-1}}
\newcommand{\Mdotunit}{M_\odot {\rm yr}^{-1}}

\shorttitle{SCORCH I}
\shortauthors{Trac, Cen, \& Mansfield}

\begin{document}

\title{SCORCH I: The Galaxy-Halo Connection in the First Billion Years}

\author{Hy Trac\altaffilmark{1}, Renyue Cen\altaffilmark{2}, and Philip Mansfield\altaffilmark{3}}
\affil{\altaffilmark{1}McWilliams Center for Cosmology, Department of Physics, Carnegie Mellon University, Pittsburgh, PA 15213 \\ 
\altaffilmark{2}Department of Astrophysical Sciences, Princeton University, Princeton, NJ 08544 \\
\altaffilmark{3}Department of Astronomy and Astrophysics, University of Chicago, Chicago, IL 60637}

\begin{abstract}
SCORCH (Simulations and Constructions of the Reionization of Cosmic Hydrogen) is a new project to study the Epoch of Reionization (EoR). In this first paper, we probe the connection between observed high-redshift galaxies and simulated dark matter halos to better understand the primary source of ionizing radiation. High-resolution N-body simulations are run to quantify the abundance of dark matter halos as a function of mass $M$, accretion rate $\dot{M}$, and redshift $z$. A new fit for the halo mass function $dn/dM$ is $\approx 20\%$ more accurate at the high-mass end. A novel approach is used to fit the halo accretion rate function $dn/d\dot{M}$ in terms of the halo mass function. Abundance matching against the observed galaxy luminosity function is used to estimate the luminosity-mass relation and the luminosity-accretion-rate relation. The inferred star formation efficiency is not monotonic with $M$ nor $\dot{M}$, but reaches a maximum value at a characteristic mass $\sim 2 \times 10^{11}\ \Msun$ and a characteristic accretion rate $\sim 6 \times 10^2\ \Mdotunit$ at $z \approx 6$. We find a universal EoR luminosity-accretion-rate relation and construct a fiducial model for the galaxy luminosity function. The Schechter parameters evolve such that $\phi_\star$ decreases, $\Mstar$ is fainter, and $\alpha$ is steeper at higher redshifts. We forecast for the upcoming James Webb Space Telescope and show that with apparent magnitude limit $m_{\rm AB} \approx 31$ (32), it can observe $\gtrsim 11$ (24) unlensed galaxies per square degree per unit redshift at least down to $\Mstar$ at $z \lesssim 13$ (14).
\end{abstract}


\section{Introduction}
\label{sec:intro}

Cosmic reionization is a frontier topic in cosmology with plenty of scientific richness for theoretical and observational explorations. The epoch of reionization (EoR) is uniquely marked by the emergence of the first luminous sources: stars, galaxies, and quasars in the first billion years. It is possibly the only time that luminous sources directly and dramatically alter the state of the entire universe, converting the cold and neutral intergalactic medium (IGM) into a warm and highly ionized one. Studying the EoR will reveal how the first generation of stars, galaxies, and black holes formed and evolved. It can also provide complementary constraints on cosmological parameters similar to studies of the cosmic microwave background (CMB). See \citet{2013fgu..book.....L} for a recent review.

Current observations suggest that reionization was already in significant progress by $z \sim 9$ and must have ended by $z \sim 6$. The photoionization and photoheating of hydrogen leaves many observable imprints. Neutral hydrogen can be detected through 21cm radiation from radio observations \citep[e.g.][]{2010Natur.468..796B, 2014ApJ...788..106P}, in absorption as the Lyman alpha forest \citep[e.g.][]{2002AJ....123.1247F, 2006AJ....132..117F}, and also through Lyman alpha scattering of high-redshift galaxies \citep[e.g.][]{2006ApJ...648....7K, 2009ApJ...706.1136O}. Free electrons will scatter CMB photons and produce patchy Thomson scattering \citep[e.g.][]{2009PhRvD..79d3003D, 2011arXiv1106.4313S, 2013ApJ...776...82N}, kinetic Sunyaev-Zel'dovich temperature anisotropy on arcminute scales \citep[e.g.][]{1970Ap&SS...7....3S, 2012ApJ...756...65Z, 2013JCAP...10..060S}, and polarization anisotropy at low multipoles \citep[e.g.][]{2003ApJS..148..161K, 2013ApJS..208...19H, 2015arXiv150201589P}.  Heating of the IGM affects the thermal broadening of the Lyman alpha forest \citep[e.g.][]{2002ApJ...567L.103T, 2009ApJ...706L.164C, 2014ApJ...788..175L}. Ongoing and upcoming observations will provide multi-wavelength constraints on the uncertain timing and duration of the EoR.

Current theory suggest that large-scale, overdense regions near radiation sources are generally reionized earlier than large-scale, underdense regions far from sources. Three main approaches are used to model the EoR \citep[e.g.][]{2011ASL.....4..228T}. Cosmological simulations combining N-body, hydro, and radiative transfer (RT) algorithms are the most accurate, but also most expensive approach to solve the coupled evolution of the dark matter, baryons, and radiation \citep[e.g][]{2000ApJ...535..530G, 2008ApJ...689L..81T, 2014ApJ...793...29G, 2015ApJS..216...16N}. RT post-processed on saved N-body or hydro simulated data can be more cost effective while capturing important features of nonlinear structure formation \citep[e.g.][]{2003MNRAS.343.1101C, 2006MNRAS.369.1625Iliev, 2007MNRAS.377.1043McQuinn, 2009MNRAS.400.1049F, 2010ApJ...724..244A}. Semi-analytical/numerical methods provide an approximate and efficient approach to explore the vast parameter space of reionization \citep[e.g.][]{2004ApJ...613....1F, 2007ApJ...654...12Z, 2010MNRAS.406.2421S, 2011MNRAS.411..955M, 2013ApJ...776...81B}. Recently, there is renewed emphasis on using radiation-hydrodynamic simulations to model the complex astrophysics of sources and sinks and using semi-numerical methods to make theoretical predictions and mock observations. These two approaches in tandem provide our best option in making forward progress in synergy with observations.

SCORCH (Simulations and Constructions of the Reionization of Cosmic Hydrogen) is a new project to study the EoR and provide useful theoretical tools and predictions to facilitate more accurate and efficient comparison between observations and theory. To build a robust theoretical framework, we need to further investigate how the distribution and properties of radiation sources and sinks affect the complex reionization process. The RadHydro code \citep{2004NewA....9..443T, 2007ApJ...671....1T} will be used to produce N-body + hydro + RT simulations with subgrid physics modeled using the latest observations and simulations. Mock observations will be constructed by mapping higher-resolution, smaller-volume radiation-hydrodynamic simulations onto lower-resolution, larger-volume N-body simulations \citep[e.g][]{2013ApJ...776...81B, 2013ApJ...776...82N, 2013ApJ...776...83B, 2014ApJ...789...31L}.

Currently, there are $\approx 1500$ galaxy candidates observed in the redshift range $6 \lesssim z \lesssim 10$ from Hubble Space Telescope (HST) observations \citep[e.g.][]{2011ApJS..197...35G, 2011ApJS..197...36K, 2011ApJS..193...27W, 2013ApJ...763L...7E, 2013ApJS..209....6I, 2014ApJ...786...57S}. The galaxy luminosity function is well fit by a Schechter function and the redshift dependence of the parameters have been quantified \citep[e.g.][]{2014ApJ...793..115B, 2014ApJ...786..108O, 2015ApJ...803...34B, 2015ApJ...810...71F}. However, it is unclear what is the physical origin for the evolution and how is it connected to the growth of structure, particularly that of dark matter halos. The abundance of dark matter halos are accurately quantified using N-body simulations \citep[e.g.][]{2001MNRAS.321..372J, 2008ApJ...688..709T}, but previous work related to the EoR \citep[e.g][]{2007MNRAS.374....2R, 2007ApJ...671.1160L} have not extensively studied atomic cooling halos ($T \gtrsim 10^4$ K, $M \gtrsim 10^8\ \Msun$) capable of hosting high-redshift galaxies. Similarly, the growth of dark matter halos have also  been measured using N-body simulations \citep[e.g][]{2002ApJ...568...52W, 2008MNRAS.386..577F, 2015MNRAS.450.1521C}, but generally for higher mass and at lower redshift.

In Paper I, we quantify the connection between observed high-redshift galaxies and simulated dark matter halos in order to better understand the abundance and evolution of the primary source of ionizing radiation. Section \ref{sec:nbody} describes the N-body code and halo finder used to simulate and locate dark matter halos. Sections \ref{sec:mass_function} and \ref{sec:mass_accretion} describe how the abundance of dark matter halos as a function of mass $M$, accretion rate $\dot{M}$, and redshift $z$ are quantified. In Sections \ref{sec:abund_match} and \ref{sec:star_formation}, we use the abundance matching technique to estimate the luminosity-mass relation and the luminosity-mass-accretion-rate relation and infer the star formation efficiency. In Sections \ref{sec:glf} and \ref{sec:jwst}, we make predictions for the high-redshift galaxy luminosity function and forecast galaxy counts for the upcoming James Webb Space Telescope (JWST). Section \ref{sec:discussion} discusses the implications of our results on cosmic reionization and Section \ref{sec:conclusions} concludes with a summary of major results. We adopt the concordance cosmological parameters: $\Omega_{\rm m} = 0.27$, $\Omega_\Lambda = 0.73$, $\Omega_{\rm b} = 0.045$, $h = 0.7$, $\sigma_8 = 0.8$, and $n_{\rm s} = 0.96$.

\newpage

\section{N-body Simulations}
\label{sec:nbody}

\begin{deluxetable*}{lcccccc}[t]
\tablewidth{\hsize}
\tabletypesize{\footnotesize}
\tablecaption{\label{tab:nbody} N-body simulation parameters}
\tablecolumns{7}
\tablehead{Name & $L$ & $m_p$ & $\epsilon$ & $z_{\rm init}$ & $M_{\rm halo,min}$ & $M_{\rm acc,min}$ \\
& ($\Mpch$) & ($\Msunh$) & ($\kpch$) & & ($\Msunh$) & ($\Msunh$)} 
\startdata
NB\_L10\_N2048 & \ 10 & $8.72 \times 10^3$ & 0.31 & $350$ & $3.49 \times 10^6\ $ & $2.18 \times 10^7\ $\\
NB\_L15\_N2048 & \ 15 & $2.94 \times 10^4$ & 0.46 & $350$ & $1.18 \times 10^7\ $ & $7.36 \times 10^7\ $\\
NB\_L25\_N2048 & \ 25 & $1.36 \times 10^5$ & 0.76 & $300$ & $5.45 \times 10^7\ $ & $3.41 \times 10^8\ $ \\
NB\_L35\_N2048 & \ 35 & $3.74 \times 10^5$ & 1.07 & $300$ & $1.50 \times 10^8\ $ & $9.35 \times 10^8\ $ \\
NB\_L50\_N2048 & \ 50 & $1.09 \times 10^6$ & 1.53 & 300 & $4.36 \times 10^8\ $ & $2.73 \times 10^9\ $ \\
NB\_L75\_N2048 & \ 75 & $3.68 \times 10^6$ & 2.29 & $300$ & $1.47 \times 10^9\ $ & $9.20 \times 10^9\ $ \\
NB\_L100\_N2048 & 100 & $8.72 \times 10^6$ & 3.05 & 250 & $3.49 \times 10^9\ $ & $2.18 \times 10^{10}$ \\
NB\_L150\_N2048 & 150 & $2.94 \times 10^7$ & 4.58 & 250 & $1.18 \times 10^{10}$ & $7.36 \times 10^{10}$ \\
NB\_L200\_N2048 & 200 & $6.98 \times 10^7$ & 6.10 & 200 & $2.79 \times 10^{10}$ & $1.74 \times 10^{11}$ \\
NB\_L300\_N2048 & 300 & $2.36 \times 10^8$ & 9.16 & 200 & $9.42 \times 10^{10}$ & $5.89 \times 10^{11}$ \\
NB\_L400\_N2048 & 400 & $5.58 \times 10^8$ & 12.2 & 150 & $2.23 \times 10^{11}$ & $1.40 \times 10^{12}$
\enddata
\tablecomments{Two realizations of each P$^3$M N-body simulation containing $2048^3$ dark matter particles are run using 2LPT initial conditions. Dark matter halos are found with a hybrid algorithm and have an average density $\bar{\rho}_{\rm halo} = 200 \bar{\rho}_m$. For calculating the halo mass and accretion rate functions, the minimum masses correspond to 400 and 2500 particles, respectively.}
\end{deluxetable*}

N-body simulations are run using a new particle-particle-particle-mesh code \citep[P$^3$M; e.g.][]{1981csup.book.....H} containing updated algorithms. The gravitational potential is decomposed into short and long-range components using a gaussian kernel following \citet{2002JApA...23..185B}. The long-range potential is computed using a particle-mesh (PM) algorithm where Poisson's equation is efficiently solved using Fast Fourier Transforms (FFTs). Particles are assigned to a mesh using the triangular-shaped-cells (TSC) scheme to reduce mesh anisotropy. The short-range force is computed for particle-particle (PP) interactions using direct summation, which is only moderately expensive at high redshifts when the large-scale structure is not highly clustered. Adaptive time stepping is performed using the kick-drift-kick scheme from \citet{2005MNRAS.364.1105S}.

Dark matter halos are found using a new hybrid finder that has two major components. First, a friends-of-friends \citep[FoF; e.g.][]{2001MNRAS.321..372J} algorithm is used to locate the peaks of halo candidates. A linking length $b = 0.08$ times the mean interparticle separation is chosen to pick out only the inner regions and prevent over-merging \citep[e.g.][]{2008MNRAS.385.2025C}. For each peak, the particle with the minimum value of the density-weighted gravitational potential $\rho\phi$ is chosen as the center.

Second, a spherical overdensity \citep[SO; e.g.][]{1994MNRAS.271..676L} algorithm is used to measure halo properties. A halo with mass $M_{\rm \Delta}$ within radius $R_{\rm \Delta}$ has an average density that is factor of $\Delta_{\rm halo}$ times the average matter density $\bar{\rho}_{\rm m}$ such that
\begin{equation}
M_\Delta = \frac{4}{3}\pi (\Delta_{\rm halo} \bar{\rho}_{\rm m}) R_\Delta^3 .
\end{equation}
Throughout the paper, $M$ and $R$ refer to $M_{200}$ and $R_{200}$, respectively. Halos are allowed to overlap, but if two candidates have centers within the larger halo's radius, one candidate is chosen as the primary halo and the other is considered a subhalo. The primary halo is the one that satisfies more of the following criteria: larger masses and more negative potentials within $R_{200}$ and $R_{500}$. Only halos with at least 400 particles ($1/\sqrt{400} = 0.05$) are counted to avoid resolution effects and ensure completeness near the low-mass end of each simulation.

The halo finder is run ``on the fly'' every 20 million cosmic years. This time interval is comparable to relevant timescales such as the halo dynamical time and the mass accretion timescale at high redshifts. In constructing the halo merger trees, every particle is tracked rather than just the most-bound particle in identified halos. We use both forward and backward linking to carry out a two-step process in progenitor-descendent matching. First, for a given progenitor of mass $M_2$ at a given time $t_2$, its descendent of mass $M_1$ from a previous time $t_1$ is the one that contributes the largest weight, which is defined to be the sum of the gravitational binding energy from particles in the progenitor that once belonged to the descendent. Second, the selected descendent may contribute particles to other halos, but the weighted contribution must be less than what it gave to the main progenitor. The mass accretion rate is calculated as
\begin{equation}
\dot{M} = \frac{M_2 - M_1}{t_2 - t_1} .
\end{equation}
Since it is more difficult to robustly measure accretion rate than mass, only halos with at least 2500 particles ($1/\sqrt{2500} = 0.02$) are used.

Table \ref{tab:nbody} lists the N-body simulations used to quantify the abundance of halos as a function of mass, accretion rate, and redshift. Box sizes in the range $10 \leq L/(\Mpch) \leq 400$ are chosen to focus on atomic cooling halos and two realizations of each box size are run to reduce sample variance. The nine largest box-size simulations are necessary to reach convergence at the lowest $M$ of interest, while the two smallest box-size simulations are for convergence at the lowest $\dot{M}$. Each simulation contains $2048^3$ dark matter particles and has a particle mass resolution $8.72 \times 10^6(L/100)^3 \ \Msunh$. The gravitational softening length is set to 1/16$^{\rm th}$ of the mean interparticle spacing. Initial conditions are generated using 2nd-order lagrangian perturbation theory \citep[2LPT;][]{1998MNRAS.299.1097S} with a linear transfer function from CAMB \citep{2002PhRvD..66j3511L}. The starting redshift $z_{\rm init}$ is chosen such that the perturbations, $\delta_{\rm rms} \sim 0.01$, are still linear.

\section{Halo mass function}
\label{sec:mass_function}

The halo mass function in differential form is defined as the comoving number density $n(>M,z)$ of halos of mass $M$ per unit mass $dM$. In extended Press-Schechter theory \citep[EPS; e.g.][]{1991ApJ...379..440B, 1993MNRAS.262..627L}, it is generally expressed as
\begin{equation}
\frac{dn}{dM} = f(\sigma)\frac{\rho_0}{M}\frac{d \ln\sigma^{-1}}{dM}, 
\label{eqn:dndM}
\end{equation}
where $\rho_0$ is the comoving average cosmic density, $\sigma$ is the rms fluctuation of the smoothed linear density field, and $f(\sigma)$ is the $\sigma$-weighted distribution of random-walk barrier-crossings. For a continuous density field, the variance of the density fluctuations is calculated as
\begin{equation}
\sigma^2(M,z) = \int P(k,z)|W(k,M)|^2 k^2 dk ,
\label{eqn:sigma}
\end{equation}
where $P(k,z)$ is the linear power spectrum extrapolated to redshift $z$ and $W(k,M)$ is the Fourier transform of the spherical tophat window function of radius $R = [M/(4/3\pi \rho_0)]^{1/3}$. Following \citet{2006ApJ...646..881W, 2008ApJ...688..709T}, the barrier-crossing distribution function is parametrized as 
\begin{equation}
f(\sigma) = A\left[1 + \left(\frac{\sigma}{b}\right)^{-a}\right]e^{-c/\sigma^2} ,
\end{equation}
where $A$ sets the overall amplitude, $a$ and $b$ set the slope and amplitude of the low-mass power law, and $c$ sets the exponential decrease scale.

Missing large-scale power due to finite box sizes are corrected using a similar procedure to \citet{2007MNRAS.374....2R}. An effective variance is calculated as
\begin{equation}
\sigma_{\rm eff}^2(M,z) = \sum_{\bf k} |\delta({\bf k},z)|^2 |W(k,M)|^2 ,
\label{eqn:sigma_discrete}
\end{equation}
where $\delta({\bf k},z)$ is the Fourier transform of the linear overdensity field extrapolated to redshift $z$. An effective mass $M_{\rm eff}$ is calculated by finding the mass in Equation \ref{eqn:sigma} which would yield a variance equal to $\sigma_{\rm eff}^2$. The effective quantities $M_{\rm eff}$ and $\sigma_{\rm eff}$ are used instead of their normal counterparts to calculate the halo mass function.

We combine all of the simulations and bin the halos to calculate the discrete halo mass function,
\begin{equation}
\frac{\Delta N}{\Delta V \Delta M} = \sum_i^N \frac{w_i}{V_i \Delta M} .
\end{equation}
Each binned halo is given a weight $w = M/M_{\rm eff}$ such that total mass in any given mass bin is conserved. Since simulations with different box sizes have different minimum masses, the effective volumes $V$ differ for the various mass bins.

\begin{figure*}[t]
\center
\includegraphics[width=0.5\textwidth]{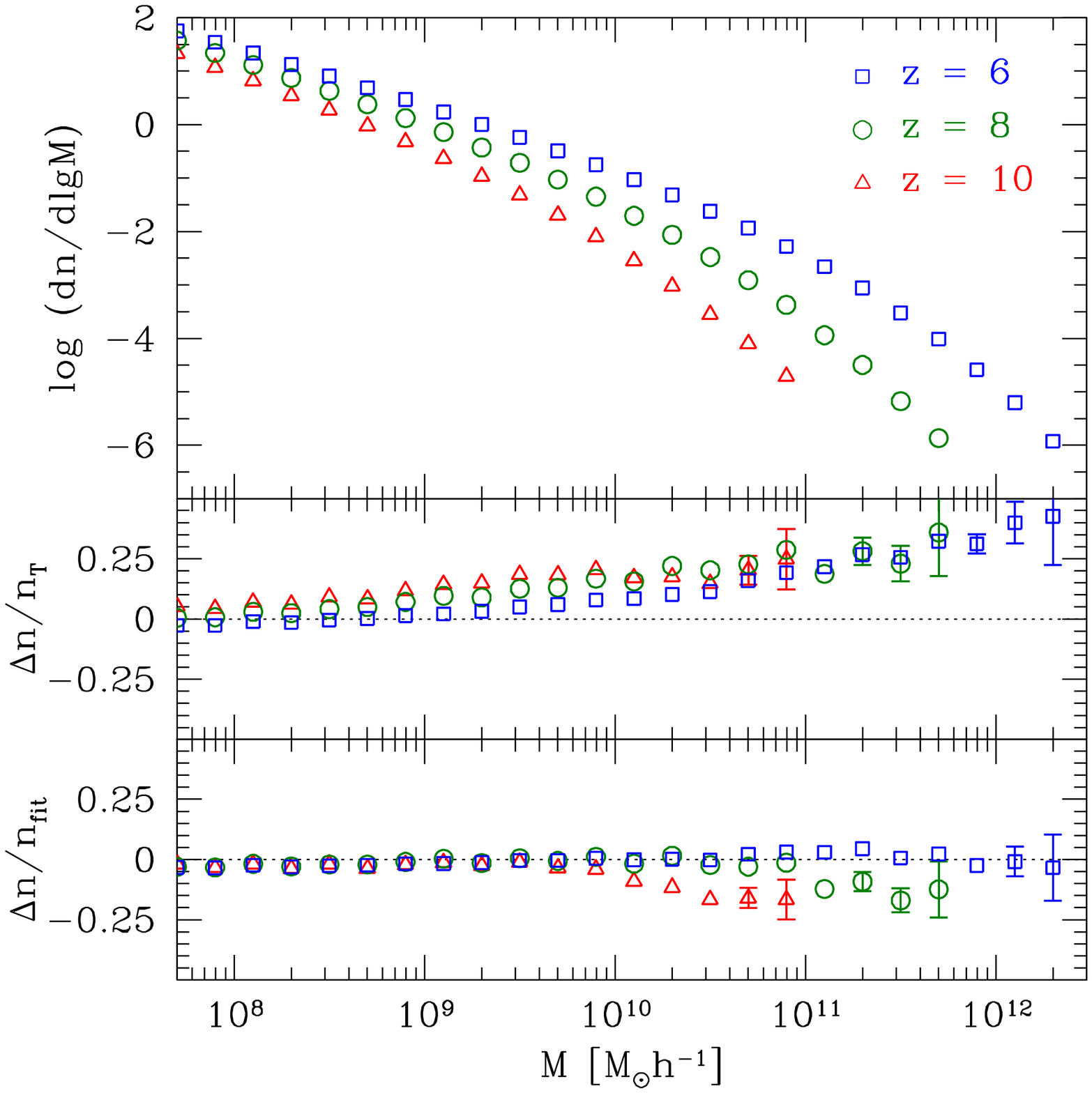}\includegraphics[width=0.5\textwidth]{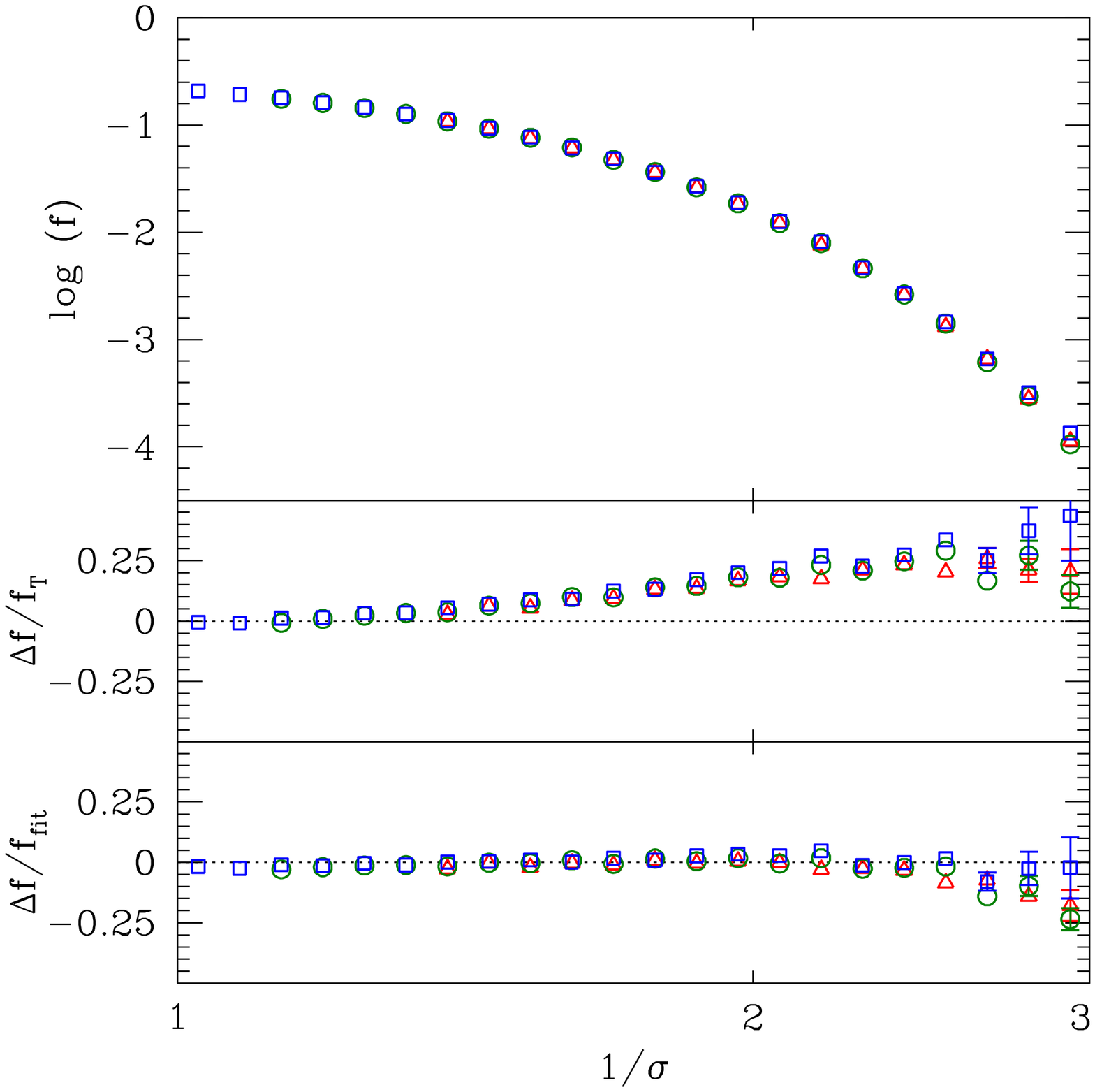}
\caption{Left: The differential halo mass function at $z=$ 6 (blue squares), 8 (green circles), and 10 (red triangles) from a series of N-body simulations (top). Comoving number densities are in units of ${\rm Mpc}^{-3}h^3$ and halo masses are in units of $\Msunh$. The simulation results are in good agreement overall with the best-fit results from \citet{2008ApJ...688..709T}, but differences of $\approx 20\%$ are seen at the high-mass end (middle). For our new fit, the residuals are typically $\lesssim 5\%$ across the mass and redshift ranges of interest (bottom). Right:  The barrier-crossing distribution function $f(\sigma)$ has a universal form for the scales and redshifts considered. The variable $\sigma^{-1}$ increases monotonically with mass. Poisson error bars are shown if they are larger than the data point symbols.}
\label{fig:mass_function}
\end{figure*}

Figure \ref{fig:mass_function} shows the differential mass functions and barrier-crossing distribution functions at $z=$ 6, 8, and 10. The mass functions have a generic shape with a low-mass power-law and a high-mass decaying tail. The function $f(\sigma)$ shows no redshift dependence after correcting for finite box-size effects. Since our box sizes are reasonably large for the mass and redshift ranges considered, the corrections are typically small, only $\sim 10\%$ at $z = 10$ and $\sim 1\%$ at $z = 6$ for moderate masses.

The middle panels of Figure \ref{fig:mass_function} show a comparison with the best-fit results from \citet{2008ApJ...688..709T}, which are calibrated for the mass range $10^{11} < M/(\Msunh) < 10^{15}$ and redshift range $0 \leq z \leq 2.5$. As suggested in their paper and through personal communication, we use a maximum redshift $z=2.5$ in their equations for the redshift evolution of the fitting parameters to obtain $A = 0.156$, $a = 1.36$, $b = 2.54$, and $c = 1.19$. The agreement is remarkably good overall considering the extrapolation in both mass and redshift. It is better at lower masses and at lower redshifts, but differences of $\approx 20\%$ are seen at the high-mass end where bright high-redshift galaxies are expected to reside. 

Since $f(\sigma)$ has a universal form for the scales and redshifts considered, we choose to fit the $z = 6$ results only because of the larger range and higher signal-to-noise. Furthermore, the parameters $a$ and $b$ are kept unchanged because the simulated halo catalogs do not sample the high-$\sigma$ (low-mass) power law portion of the function. The best-fit barrier-crossing distribution function is
\begin{equation}
f(\sigma) = 0.150\left[1 + \left(\frac{\sigma}{2.54}\right)^{-1.36}\right]e^{-1.14/\sigma^2} .
\label{eqn:fsigma}
\end{equation}
The bottom panels of Figure \ref{fig:mass_function} show that the residuals for the new fit are $\lesssim 5\%$ across the mass and redshift ranges of interest. The larger deficits at the highest mass and particularly at higher redshifts arise because we cannot perfectly correct for the missing large-scale power in finite simulation boxes. In the exponential tail of the halo mass function, a small mass change can lead to a relatively large change in the number density. While we have started the simulations at high enough initial redshifts such that the second-order displacement corrections are themselves small, starting at even high redshifts may reduce the mass deficits.

The halo mass function can be calculated for any cosmological model given the linear power spectrum. For our set of cosmological parameters, the tophat variance can be fit to $\lesssim 2\%$ accuracy for $10^7 < M/(\Msunh) < 10^{13}$ with
\begin{equation}
\sigma^{-1}(z)  = D^{-1}(z)\left[0.11 + 0.077\left(\frac{M}{10^8\ \Msunh}\right)^{0.18}\right] ,
\end{equation}
where $D(z)$ is the linear growth factor. Note that a double power law plus a constant term can accurately fit the entire mass range spanning mini-halos ($M \sim 10^5\ \Msunh$) to massive clusters ($M \sim 10^{15}\ \Msunh$).

\section{Mass Accretion Rates}
\label{sec:mass_accretion}

The halo accretion rate function in differential form is defined as the comoving number density $n(>\dot{M},z)$ of halos with accretion rate $\dot{M}$ per unit accretion rate $d\dot{M}$. In principle, its functional form in terms of accretion rate and redshift can be motivated with EPS theory. Alternatively, we take a novel approach and express it as
\begin{equation}
\frac{dn}{d\dot{M}} = \frac{dn}{dM}\frac{dM}{d\dot{M}},
\label{eqn:dndmdot}
\end{equation}
where $dn/dM$ is the differential mass function (Equation \ref{eqn:dndM}). The advantage of using this functional form is that the redshift dependence of the halo mass function is already well known. To use Equation \ref{eqn:dndmdot}, we need to understand how mass and accretion rate are related and be able to calculate the derivative $dM/d\dot{M}$.

\begin{figure*}[t]
\center
\includegraphics[width=0.5\textwidth]{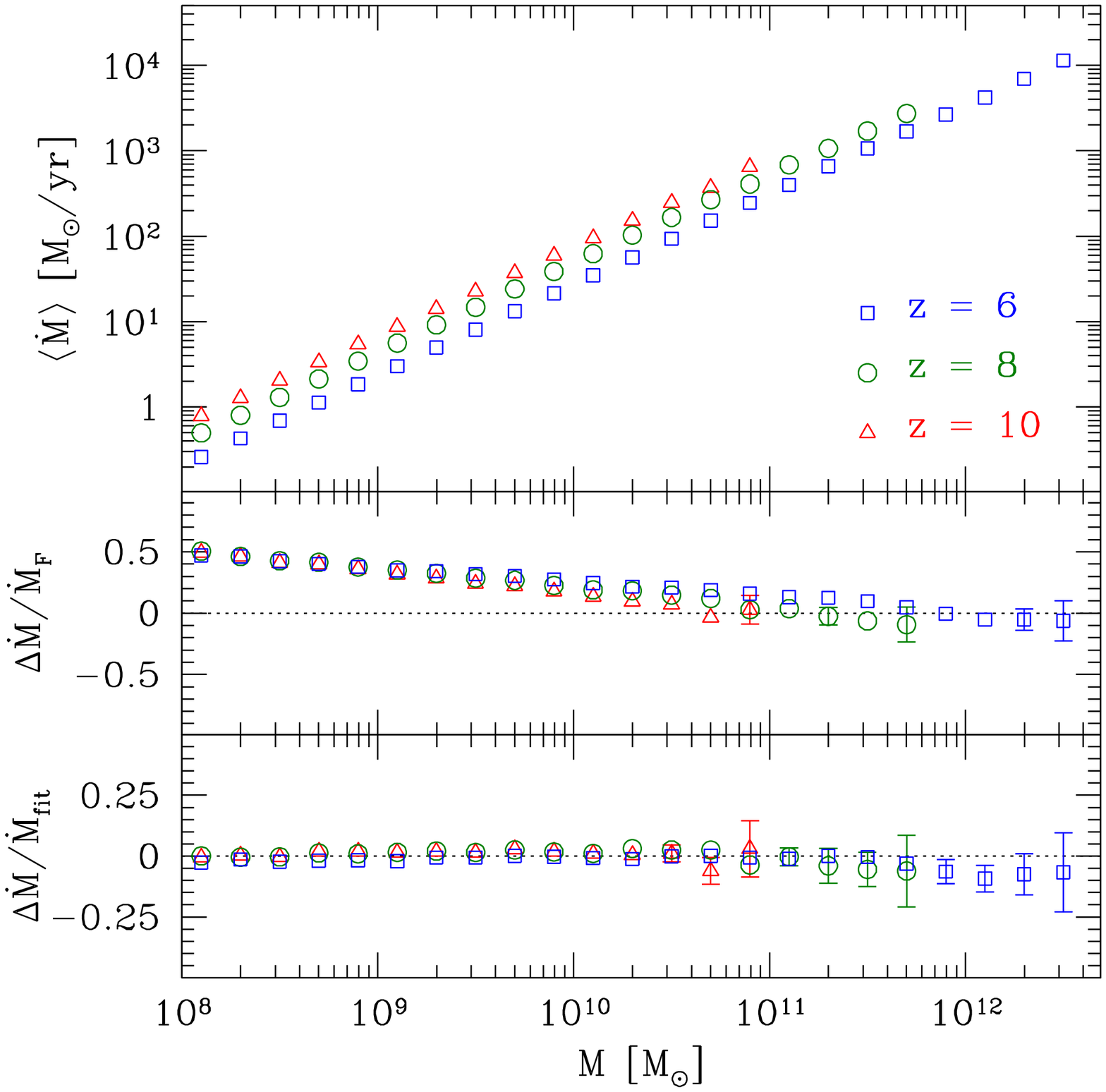}\includegraphics[width=0.5\textwidth]{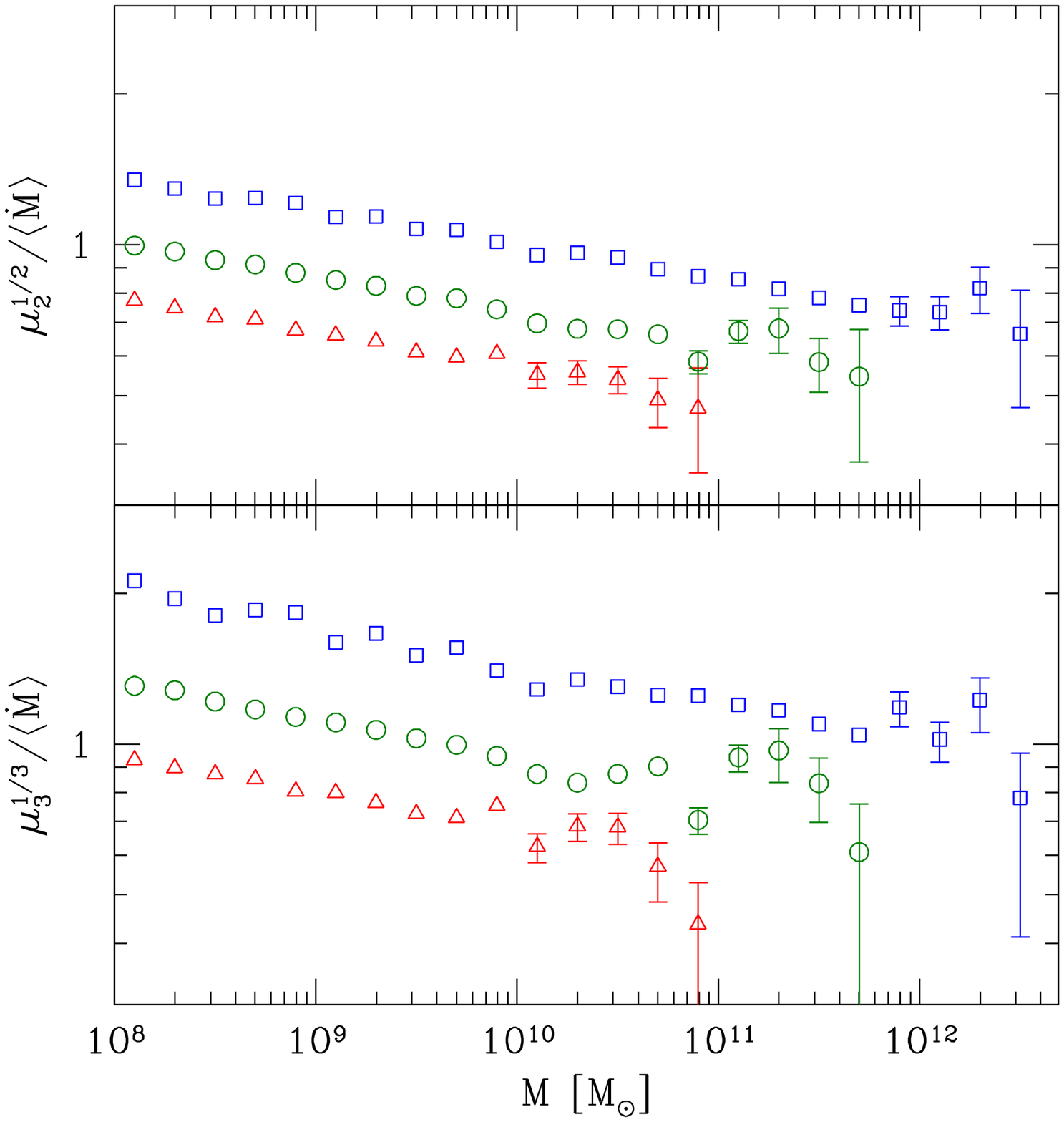}
\caption{Left: The average mass accretion rates at $z=$ 6 (blue squares), 8 (green circles), and 10 (red triangles) are accurately fit by a $\langle\dot{M}\rangle \propto M^{1.06}(1+z)^{2.5}$ relation (top). The SO halo rates are generally higher than the FoF halo rates from \citet{2010MNRAS.406.2267F}, but there is better agreement in the overlap mass range $M > 10^{10}\, M_\odot$ (middle). For our new fit, the residuals are typically $\lesssim 5\%$ across the mass and redshift ranges of interest (bottom). Right: The variance and skewness also have power-law relations with mass, but with slightly shallower slopes and weaker redshift dependences. The statistical uncertainties in the binned mean values are shown if the error bars are larger than the data point symbols.}
\label{fig:mdot}
\end{figure*}

Figure \ref{fig:mdot} shows how the mass accretion rate depends on mass at $z=$ 6, 8, and 10. The average accretion rates for the mass range $10^8 < M < 10^{13} $ and redshift range $6 \leq z \leq 10$ are well fit by
\begin{equation}
\langle\dot{M}\rangle = 0.21\, \Mdotunit \left(\frac{M}{10^8\, M_\odot}\right)^{1.06}\left(\frac{1+z}{7}\right)^{2.5} .
\label{eqn:mdot_avg}
\end{equation}
The residuals are typically $\lesssim 5\%$ across the mass and redshift ranges of interest. The larger deficits at the highest masses and particularly at higher redshifts are related to those seen in the halo mass function in Figure \ref{fig:mass_function} since the accretion rates are also affected by the missing large-scale power in finite simulation boxes.

Our mass power-law slope of 1.06 is slightly shallower than the value of $1.1$ from \citet{2010MNRAS.406.2267F} and slightly steeper than the linear mass dependence from \citet{2015MNRAS.450.1514C, 2015MNRAS.450.1521C}. We find similar results for $M \gtrsim 10^{10}\ \Msun$ where our mass range overlaps with theirs, but our accretion rates become relatively larger (smaller) at lower mass because of the shallower (steeper) slope. We all find the same redshift power-law slope of 2.5 at high $z$. \citet{2010MNRAS.406.2267F} measure accretion rates from N-body simulations, but they use a FoF halo finder instead. Since the ratio of $M_{200}$ and $M_{\rm FoF}$ is neither constant with mass nor redshift, it is not surprising to find differences between $\dot{M}_{200}$ and $\dot{M}_{\rm FoF}$. Their fit is done at low redshift $z \sim 0$ and it is not clear how well their best fit compares to their own data for $6 \leq z \leq 10$. \citet{2015MNRAS.450.1514C, 2015MNRAS.450.1521C} use EPS theory and N-body simulations to derive and measure the mass accretion history, respectively. They find a linear dependence on mass from their analytical work and assume this same mass dependence when fitting the numerical results. Thus, it is not clear if the latter would have had a slightly higher mass power-law slope. Nonetheless, there is overall good agreement between all of the results given the differences in details.

The distribution of mass accretion rate at any given mass is positively skewed. The best-fit relations for the variance $\mu_2 \equiv  \langle(\dot{M} - \langle\dot{M}\rangle)^2\rangle$ and skewness $\mu_3 \equiv  \langle(\dot{M} - \langle\dot{M}\rangle)^3\rangle$ are
\begin{align}
\label{eqn:mdot_var}
\mu_2^{1/2} & = 0.28\, \Mdotunit \left(\frac{M}{10^8\, M_\odot}\right)^{1.0}\left(\frac{1+z}{7}\right)^{1.1}, \\
\label{eqn:mdot_ske}
\mu_3^{1/3} & =  0.42\, \Mdotunit \left(\frac{M}{10^8\, M_\odot}\right)^{1.0}\left(\frac{1+z}{7}\right)^{0.6} .
\end{align}
Equations \ref{eqn:mdot_avg}-\ref{eqn:mdot_ske} can be used with, for example, an Edgeworth expansion to model the probability distribution function of accretion rates at a given mass and redshift. Mergers are responsible for the positive tail of the distribution. Thus, the similarity in mass dependence and difference in redshift evolution between $\langle \dot{M} \rangle$, $\mu_2^{1/2}$, and $\mu_3^{1/3}$ reflect those between the average accretion rate and the average merger rate. See \citet{2010MNRAS.406.2267F} for recent work on halo merger rates.

\begin{figure}[t]
\center
\includegraphics[width=0.5\textwidth]{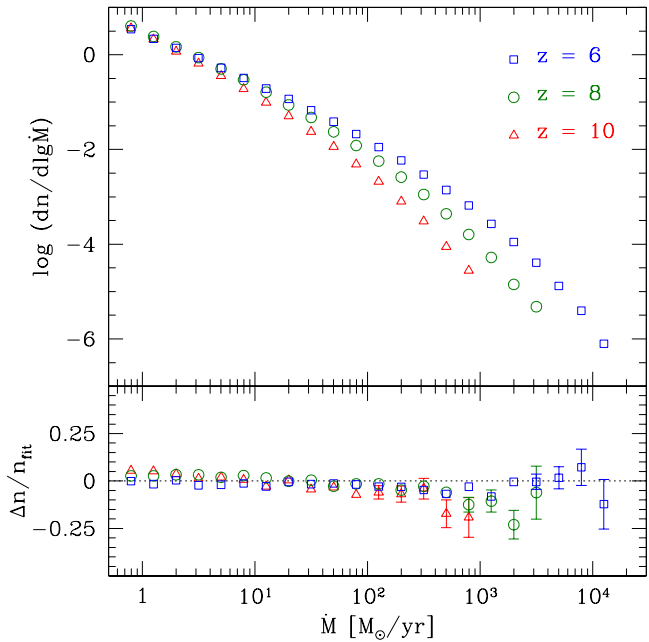}
\caption{The differential halo accretion rate function at $z=$ 6 (blue squares), 8 (green circles), and 10 (red triangles) from a series of N-body simulations (top). Comoving number densities are in units of Mpc$^{-3}$ and mass accretion rates are in units of $\Mdotunit$. The halo mass function and the halo accretion rate function are similar in shape, with each having a low-end power law and a high-end exponential decline. The latter can be predicted from the former using Equations \ref{eqn:dndmdot} and \ref{eqn:mass_mdot}. The residuals are typically $\lesssim 10\%$ for the mass and redshift ranges of interest (bottom). Poisson error bars are shown if they are larger than the data point symbols.}
\label{fig:dndlgmdot}
\end{figure}

Figure \ref{fig:dndlgmdot} shows the abundance of dark matter halos as a function of mass accretion rate at $z=$ 6, 8, and 10. The halo accretion rate function and the halo mass function are similar in shape, with each having a low-end power law and a high-end exponential decline. We can model the halo accretion rate function using Equation \ref{eqn:dndmdot} and by combining the halo mass function with the best-fit mediating mass,
\begin{equation}
M = 4.2 \times 10^8\Msun \left(\frac{\dot{M}}{\Mdotunit}\right)^{0.91}\left(\frac{1+z}{7}\right)^{-2.4} .
\label{eqn:mass_mdot}
\end{equation}
The bottom panel of Figure \ref{fig:dndlgmdot} shows that the residuals for the best fit are typically $\lesssim 10\%$ for the mass and redshift ranges of interest. The larger deficits at the highest $\dot{M}$ and particularly at higher redshifts are related to those seen in the halo mass function in Figure \ref{fig:mass_function} since the accretion rates are also affected by the missing large-scale power in finite simulation boxes. Note that Equation \ref{eqn:mass_mdot} is not simply obtained from Equation \ref{eqn:mdot_avg} because the distribution of accretion rate at a given mass is skewed. However, they both have the convenient property of separable power-law dependences on $M$, $\dot{M}$, and $z$. 

The halo accretion rate function does not have a simple redshift dependence according to Equation \ref{eqn:dndmdot}. For a given accretion rate, both the mediating mass and its derivative increase at lower redshifts. This is a consequence of accretion slowing down as seen in Figure \ref{fig:mdot}. In contrast, the halo mass function at a higher mass decreases in amplitude because more massive halos are rarer in hierarchical structure formation. We expect the redshift dependence in Equations \ref{eqn:mdot_avg}-\ref{eqn:mass_mdot} to hold for redshifts $z \gtrsim 6$ relevant to the EoR. However, our single redshift power law scaling is not appropriate at lower redshifts. \citet{2009MNRAS.398.1858M} have shown that a more complex scaling is required to parametrize the entire mass accretion history. More work is required to examine the validity and accuracy of extrapolating our results to lower redshifts.

\newpage

\section{Abundance matching}
\label{sec:abund_match}

\begin{figure*}[t]
\center
\includegraphics[width=0.5\textwidth]{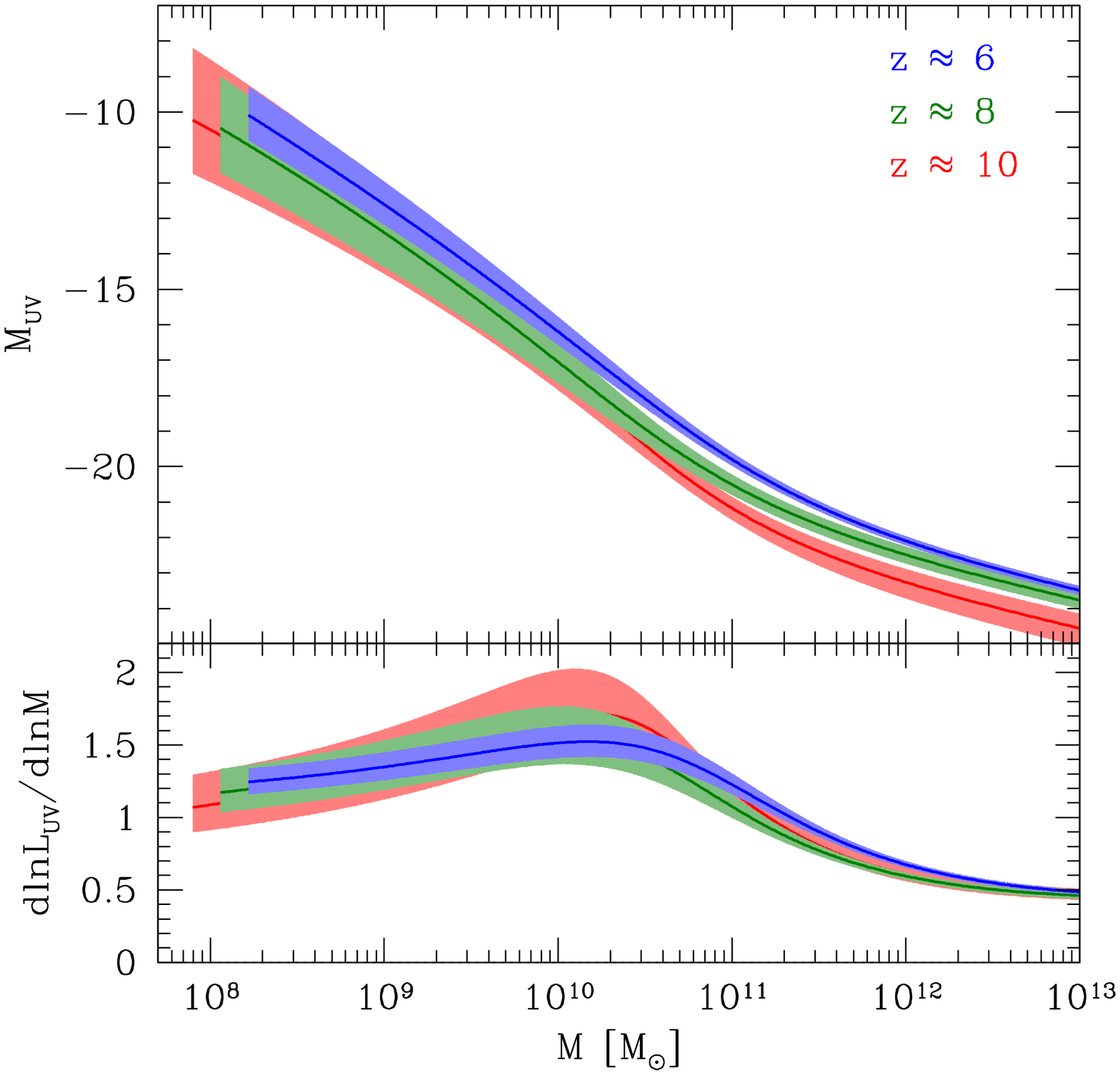}\includegraphics[width=0.5\textwidth]{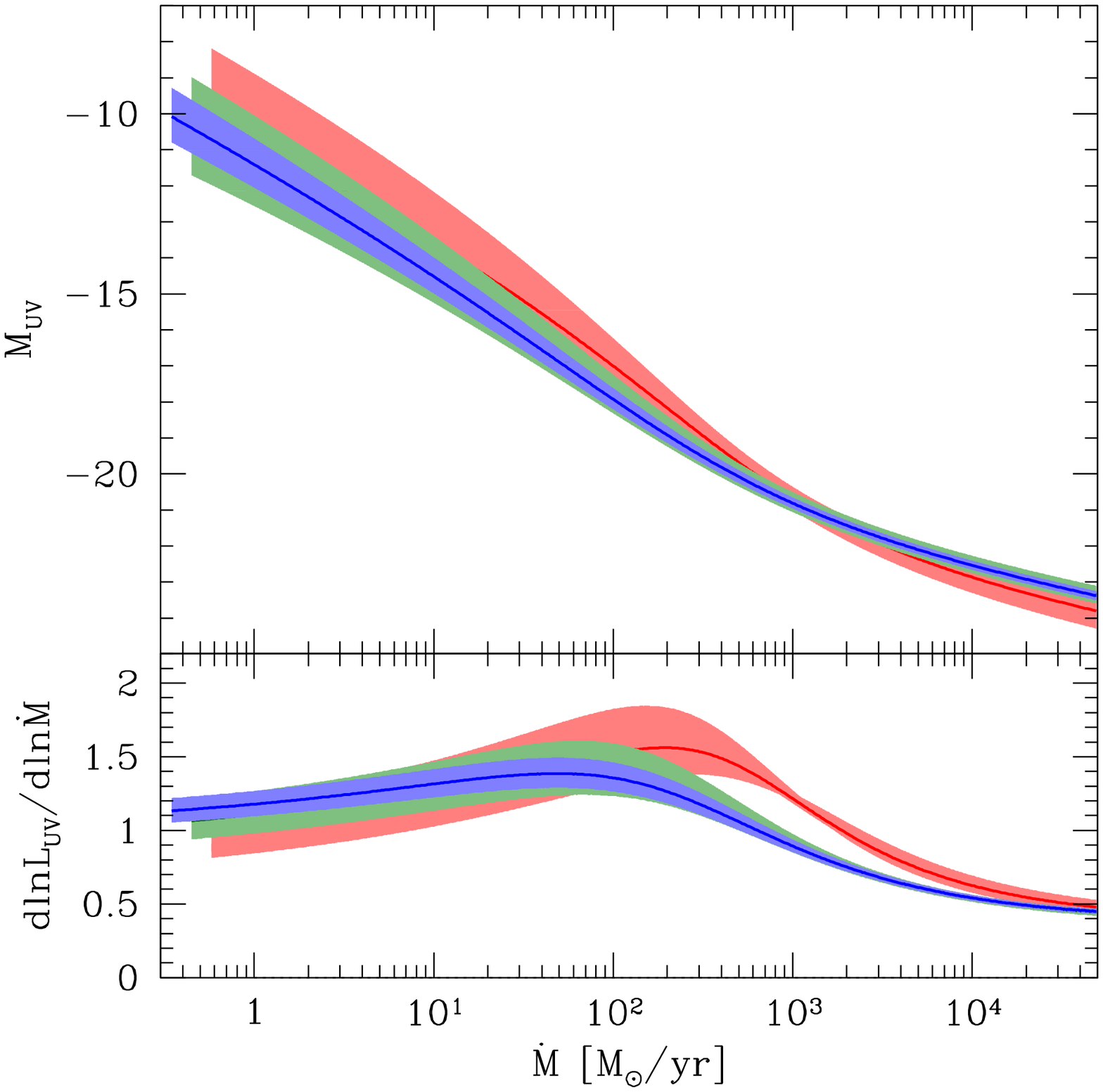}
\caption{Left: The UV magnitude as a function of halo mass for $z\approx$ 6 (blue), 8 (green), and 10 (red) from abundance matching $\Luv$ and $M$. The shaded curves are illustrative of the $1\sigma$ (68\%) uncertainty in the currently-observed galaxy luminosity functions and are truncated at the star formation limit. Right: The UV magnitude as a function of mass accretion rate from abundance matching $\Luv$ and $\dot{M}$. The luminosity-accretion relation is more consistent with no redshift evolution than the luminosity-mass relation for the redshift range $6 \lesssim z \lesssim 10$. The luminosity monotonically increases with mass and accretion rate, but the effective power-law slopes suggest that the relations should be fit with, for example, triple power law functions (bottom).}
\label{fig:Muv}
\end{figure*}

Abundance matching is performed by equating the number density of galaxies to the number density of halos in differential form,
\begin{equation}
\frac{dn_{\rm gal}}{d\Luv} = \frac{dn_{\rm halo}}{dX}\frac{dX}{d\Luv} ,
\label{eqn:dndL}
\end{equation}
or in cumulative form,
\begin{equation}
n_{\rm gal}(>\Luv,z) = n_{\rm halo}(>X,z) .
\end{equation}
If $X$ is taken to be the halo mass $M$, then we infer the luminosity-mass relation $\Luv(M,z)$. This procedure does not account for scatter in the mass-to-light ratio and the relation should be interpreted as an average luminosity for a given mass. If $X$ is taken to be the mass accretion rate $\dot{M}$, then we infer the luminosity-accretion rate relation $\Luv(\dot{M},z)$. This procedure accounts for the episodic nature of star formation and scatter in the mass-to-light ratio. Furthermore, it is more logical since the star formation rate should be more highly correlated with the halo growth rate rather than halo mass.

The galaxy luminosity function is generally parametrized with a Schechter function in luminosity form,
\begin{equation}
\phi(\Luv) = \phi_\star \left(\frac{\Luv}{L_\star}\right)^\alpha \exp\left(-\frac{\Luv}{L_\star}\right) ,
\label{eqn:phi_L}
\end{equation}
or in magnitude form,
\begin{align}
\phi(\Muv) = (0.4\ln 10) \phi_\star & \left[10^{0.4(M_\star - \Muv)}\right]^{\alpha+1} \nonumber \\
\times & \exp\left[10^{0.4(M_\star - \Muv)}\right] ,
\label{eqn:phi_Muv}
\end{align}
where $\phi_\star$ is an overall normalization, $L_\star$ is a characteristic luminosity, $M_\star$ is a characteristic magnitude, and $\alpha$ is the slope of the faint-end power law. The conversion between UV magnitude and luminosity is given by the standard AB relation,
\begin{equation}
M_{\rm UV} = -2.5\log\left(\frac{\Luv}{4.487 \times 10^{20}\ {\rm erg\, s^{-1}\, Hz^{-1}}}\right) .
\label{eqn:Muv_Luv}
\end{equation}
The cumulative galaxy luminosity function is obtained by integrating Equation \ref{eqn:phi_L} analytically to get the incomplete gamma function $\phi_\star L_\star \Gamma(1+\alpha,\Luv/L_\star)$. The cumulative halo mass function and mass accretion rate function do not have simple analytical forms, but are easily obtained by numerical integration.

\citet{2015ApJ...803...34B} have made the most recent measurements of the high-redshift UV luminosity functions. For three redshift samples with $\langle z \rangle \approx 5.9$, 7.9, and 10.4, there are 867, 217, and 6 galaxy candidates observed, respectively. For reference, their best-fit Schechter parameters are:
\begin{align}
\langle z \rangle \approx \ & 5.9, & \phi_\star = 5.0 \times 10^{-4},\ \Mstar = -20.94,\ \alpha = -1.87 \nonumber \\
\langle z \rangle \approx \ & 7.9, & \phi_\star = 2.1 \times 10^{-4},\ \Mstar = -20.63,\ \alpha = -2.02 \nonumber \\
\langle z \rangle \approx \ & 10.4, & \phi_\star = 8.0 \times 10^{-5},\ \Mstar = -20.92,\ \alpha = -2.27 \nonumber \\
z \rightarrow\ & 10.4, & \phi_\star = 3.0 \times 10^{-4},\ \Mstar = -20.92,\ \alpha = -2.27 \nonumber
\end{align}
where $\phi_\star$ has units of comoving Mpc$^{-3}$. Note that the last entry above is not a best fit to the galaxy counts, but an extrapolation of the luminosity functions from lower redshifts. They allowed the parameters $\log \phi_\star$, $M_\star$, and $\alpha$ to vary linearly with $z$ and fitted the galaxy counts in the redshift range $4 \lesssim z \lesssim 8$ in order to estimate the results for higher redshifts. Using the derived redshift dependence, they were able to fix $\Mstar$ and $\alpha$ and then fit for $\phi_\star$ at $\langle z \rangle \approx 10.4$. We note that $\Mstar = -20.92$ at $z \approx 10$ is rather bright considering that the best-fit characteristic magnitude changed from -20.94 to -20.63 for $z \approx 6$ to 8. Furthermore, the extrapolation from a different redshift range $5 \lesssim z \lesssim 8$ yields $\Mstar \approx -20.22$ \citep{2015arXiv150601035B}, which is more in line with the best-fit values for $z \approx 6-8$.

To estimate upper and lower bounds on a given galaxy luminosity function, we vary the three Schechter parameters using their uncertainties, add the weighted Schechter functions to the $\Muv$-$\phi$ grid, and calculate the 68\% confidence level. For the weight, we use a trivariate gaussian likelihood that assumes uncorrelated errors. However, there is degeneracy between the parameters and their errors are correlated. To prevent overestimating the confidence region, we reduce the errors in the Schechter parameters to 2/3 of their original values. This procedure allows us to match the $1\sigma$ errors in the binned measurements of the galaxy luminosity function. A more rigorous statistical analysis would require using the full likelihood for the galaxy luminosity function fit, which is not available.

Figure \ref{fig:Muv} shows how the UV luminosity and magnitude depend on mass and accretion rate at $z \approx 6$, 8, and 10 (i.e.~5.9, 7.9, and 10.4). The shaded curves are illustrative of the $1\sigma$ (68\%) uncertainty in the current galaxy luminosity function. Our error analysis is able to capture the trend of increasing uncertainty at higher redshifts. Neither the luminosity-mass relation nor the luminosity-accretion-rate relation is represented by a single power law, but appears to be composed of several power law portions.

To qualitatively understand the shape of the luminosity-mass and luminosity-accretion-rate relations, we also plot their effective power-law slopes in the bottom panels of Figure \ref{fig:Muv}. The cumulative galaxy and halo number densities can be written in effective power form as $n_{\rm gal}(L) \propto L^{1+\alpha_{\rm eff}}$, $n_{\rm halo}(M) \propto M^{1+\beta_{\rm eff}}$, and $n_{\rm halo}(\dot{M}) \propto \dot{M}^{1+\gamma_{\rm eff}}$. The luminosity-mass and luminosity-accretion-rate relations can be written as $\Luv \propto M^{\delta_{\rm eff}}$ and $\Luv \propto \dot{M}^{\epsilon_{\rm eff}}$. From Equation \ref{eqn:dndL}, the slopes are related through
\begin{gather}
\delta_{\rm eff} = \frac{d\ln \Luv}{d\ln M} = \frac{d\ln n_{\rm halo}}{d\ln M} {\Big /} \frac{d\ln n_{\rm gal}}{d\ln \Luv} = \frac{1+\beta_{\rm eff}}{1+\alpha_{\rm eff}} , \\[5pt]
\epsilon_{\rm eff} = \frac{d\ln \Luv}{d\ln\dot{M}} = \frac{d\ln n_{\rm halo}}{d\ln\dot{M}} {\Big /} \frac{d\ln n_{\rm gal}}{d\ln \Luv} = \frac{1+\gamma_{\rm eff}}{1+\alpha_{\rm eff}} .
\end{gather}
At low $\Luv$, $M$, and $\dot{M}$, where the galaxy luminosity function and halo mass and accretion rate functions are approximately power laws ($\alpha_{\rm eff}\approx \alpha$, $\beta_{\rm eff} \approx -2$, $\gamma_{\rm eff} \approx -2$), the effective power-law slopes of the luminosity-mass and luminosity-accretion-rate relations are approximately constant $\delta_{\rm eff} \approx \epsilon_{\rm eff} \approx -1/(1+\alpha)$. At intermediate scales, the halo mass and accretion rate functions deviate from their power law forms before the galaxy luminosity function does, both $\beta_{\rm eff}$ and $\gamma_{\rm eff}$ become more negative, and both $\delta_{\rm eff}$ and $\epsilon_{\rm eff}$ increase. As the galaxy luminosity function deviates from its power law form, $\alpha_{\rm eff}$ becomes more negative, and both $\delta_{\rm eff}$ and $\epsilon_{\rm eff}$ decrease again. At the very bright and massive end, both the galaxy and halo number densities are exponentially decreasing, but such that $\alpha_{\rm eff} < \beta_{\rm eff}$ and $\alpha_{\rm eff} < \gamma_{\rm eff}$, resulting in $\delta_{\rm eff}$ and $\epsilon_{\rm eff}$ being approximately constant. The effective power-law slopes suggest that both relations can be fit with, for example, a triple power law,
\begin{equation}
\Luv = L_0 \left(\frac{X}{X_{\rm a}}\right)^a \left(1+\frac{X}{X_{\rm b}}\right)^{b-a} \left(1+\frac{X}{X_{\rm c}}\right)^{c-b} ,
\end{equation}
where $L_0$ is an overall amplitude, $a$, $b$, and $c$ are three power-law slopes, and $X_{\rm a}$, $X_{\rm b}$, and $X_{\rm c}$ are three characteristic scales.

For the luminosity-mass relation, brighter galaxies are found in more massive halos at any given redshift as expected from the abundance matching procedure. At any given mass, brighter galaxies are found at higher redshifts and this is a consequence of the relative evolution of the galaxy luminosity and halo mass functions. Our results are similar to \citet{2012MNRAS.423..862K}, where they performed abundance matching using older fits to the galaxy luminosity function and halo mass function.

For the luminosity-accretion-rate relation, brighter galaxies are found in halos with larger accretion rates at any given redshift by construction. The $z \approx 6 - 8$ results are remarkably very similar for the entire accretion rate of interest. In fact, the luminosity-accretion-rate relation is consistent with no evolution for the redshift range $6 \lesssim z \lesssim 10$ when taking into account the current uncertainties in the galaxy luminosity functions. In Section \ref{sec:glf}, we show that our results at $z \approx 10$ are consistent with the binned measurements of the galaxy luminosity function \citep[e.g.][]{2014ApJ...786..108O, 2015ApJ...803...34B}.

The luminosity-accretion-rate relation at $z \approx 6$ can be fit with a triple power law,
\begin{align}
\Luv = 1.5 \times 10^{25}\ {\rm erg\, s^{-1}\, Hz^{-1}} & \left(\frac{\dot{M}}{\Mdotunit}\right)^{1.15}\nonumber \\
\times & \left(1+\frac{\dot{M}}{3.5\, \Mdotunit}\right)^{0.25} \nonumber \\
\times & \left(1+\frac{\dot{M}}{1000\, \Mdotunit}\right)^{-0.95},
\label{eqn:Luv_Mdot}
\end{align}
\begin{align}
\Muv = -11.3 & - 2.88\log\left(\frac{\dot{M}}{\Mdotunit}\right) \nonumber \\
& - 0.63\log\left(1+\frac{\dot{M}}{3.5 \, \Mdotunit}\right) \nonumber \\
& + 2.38\log\left(1+\frac{\dot{M}}{1000\, \Mdotunit}\right) .
\label{eqn:Muv_Mdot}
\end{align}
Because of the degeneracy between the triple power law parameters, we choose to fix the low, middle, and high-$\dot{M}$ slopes at 1.15, 1.40, and 0.45, respectively based on Figure \ref{fig:Muv}. There are still degeneracies between the amplitude and the characteristic accretion rates. The residuals for the fit are $\lesssim 5\%$ across the accretion rate range of interest. If we assume no evolution in the luminosity-accretion-rate relation, then Equations \ref{eqn:Luv_Mdot} and \ref{eqn:Muv_Mdot} can be used as a universal EoR template to construct a fiducial model for the evolution of the galaxy luminosity function as discussed in Section \ref{sec:glf}.

\section{Star Formation Efficiency}
\label{sec:star_formation}

\begin{figure*}[t]
\center
\includegraphics[width=0.5\textwidth]{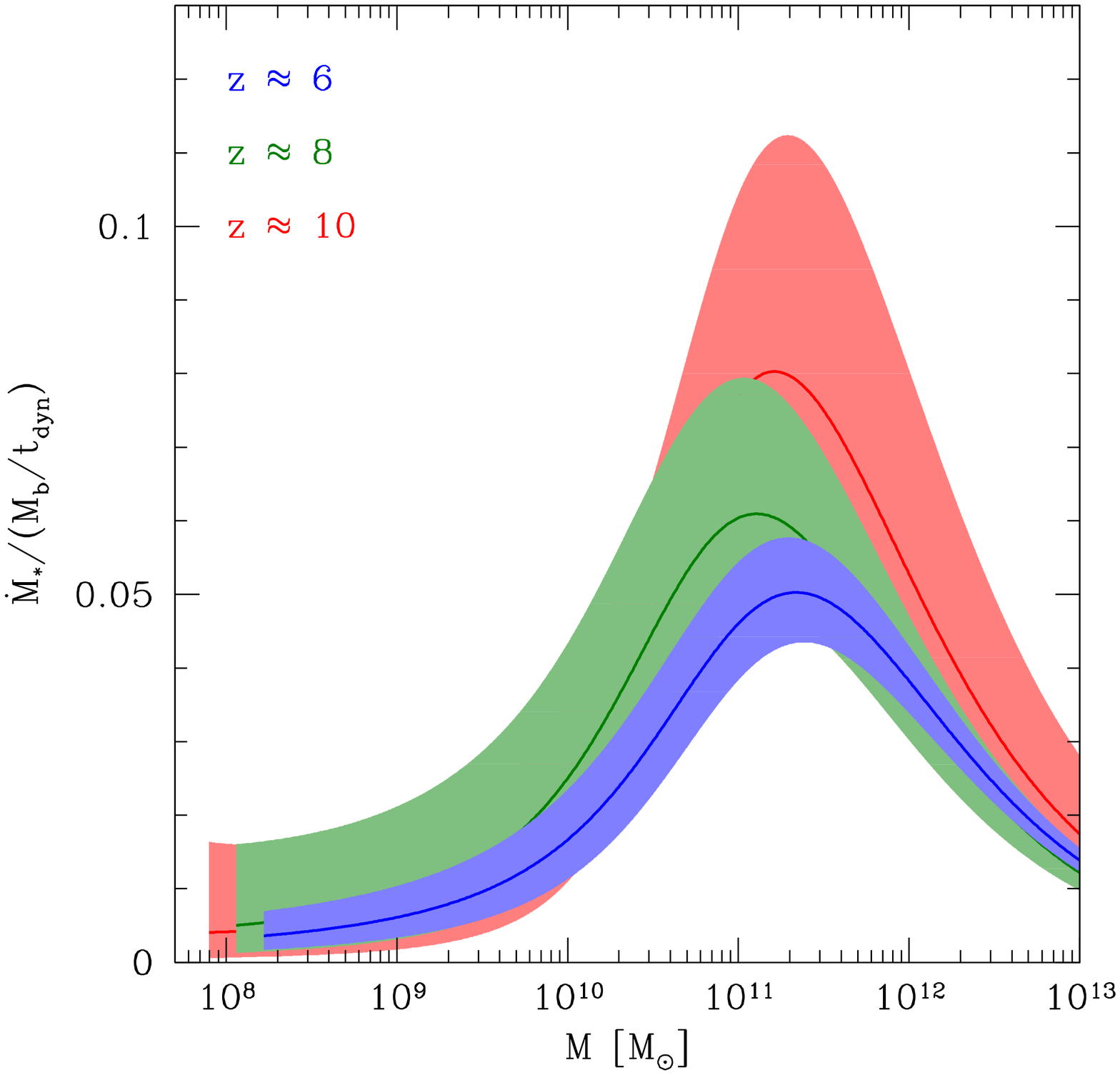}\includegraphics[width=0.5\textwidth]{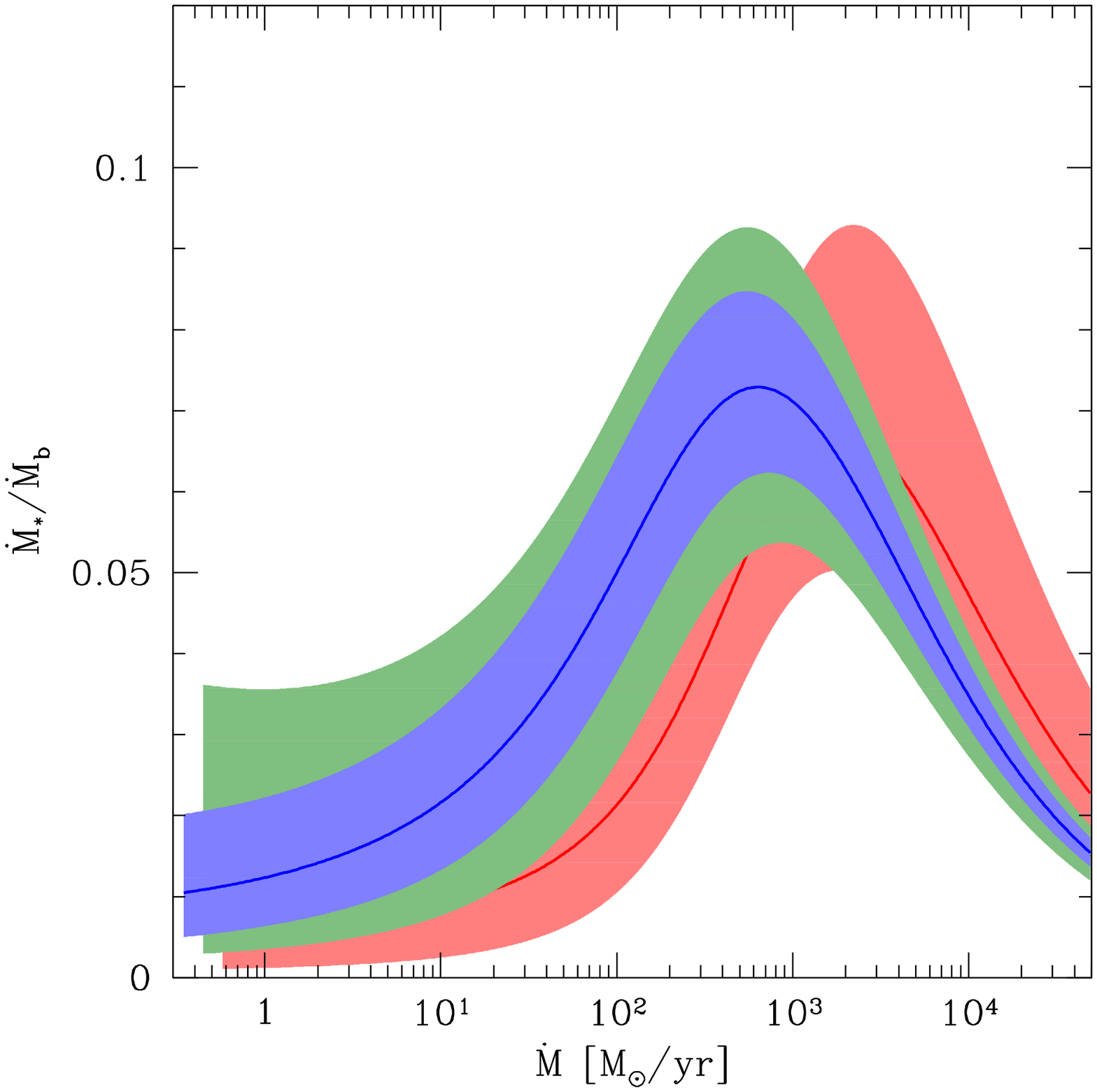}
\caption{Left: The star formation efficiency as a function of halo mass for $z \approx 6$ (blue), 8 (green), and 10 (red) from abundance matching $\Luv$ and $M$. The shaded curves are illustrative of the $1\sigma$ (68\%) uncertainty in the currently-observed galaxy luminosity functions. Right: The star formation efficiency as a function of accretion rate from abundance matching $\Luv$ and $\dot{M}$. The results are consistent with no evolution for the redshift range $6 \lesssim z \lesssim 10$.}
\label{fig:sfe}
\end{figure*}

To better understand the high-redshift galaxy-halo connection from abundance matching, we examine the implications of the inferred luminosity-mass and luminosity-accretion rate relations on the star formation rate and efficiency. The star formation rate $\dot{M}_{\rm s}$ can be estimated from the UV luminosity using the standard relation from \citet{1998ApJ...498..106M}:
\begin{equation}
\dot{M}_{\rm s} = \left(\frac{\Luv}{8 \times 10^{27}\ {\rm erg\, s^{-1} Hz^{-1}}}\right)\Mdotunit , 
\end{equation}
assuming a Salpeter initial mass function (IMF) truncated at 0.1 $\Msun$ and 125 $\Msun$. The normalization is uncertain and depends on the stellar IMF and formation history, but we adopt this commonly used relation to allow a wider comparison with other work.

We calculate the star formation efficiency in different ways for the two abundance matching results. In the case of the luminosity-accretion-rate relation, the star formation efficiency is defined as
\begin{equation}
\varepsilon_{\rm \dot{M}} \equiv \frac{\dot{M}_{\rm s}}{\dot{M}_{\rm b}} ,
\end{equation}
where the baryonic mass accretion rate,
\begin{equation}
\dot{M}_{\rm b} = \frac{\Omega_{\rm b}}{\Omega_{\rm m}}\dot{M} \propto (1+z)^{2.5},
\end{equation}
is calculated assuming the cosmic baryon fraction. In the case of the luminosity-mass relation, we assume no information about accretion rates and define the efficiency as
\begin{equation}
\varepsilon_{\rm M} \equiv \frac{\dot{M}_{\rm s}}{M_{\rm b}/t_{\rm dyn}} ,
\end{equation}
where the baryonic mass consumption rate,
\begin{equation}
\frac{M_{\rm b}}{t_{\rm dyn}} = \left(\frac{\Omega_b}{\Omega_{\rm m}}M \right)\left(\frac{32 G \bar{\rho}_{\rm halo}}{3}\right)^{1/2} \propto (1+z)^{1.5} .
\label{eqn:Mb_dot}
\end{equation}
depends on the redshift-dependent halo dynamical time. Equation \ref{eqn:Mb_dot} can be considered the 3-dimensional analog of the observed Schmidt-Kennicutt relation and is often used in semi-analytical models and cosmological simulations of galaxy formation. The star formation efficiency is expected to range from $\sim 0.01$ to $\sim 0.1$, but keep in mind that the normalization is uncertain.

We impose a star formation limit based on the criterion that galaxies form in dark matter halos where the gas cools efficiently through atomic transitions. For a minimum virial temperature $T_{\rm min} = 10^4$ K, the minimum mass for a halo to host a galaxy is defined as
\begin{align}
& M_{\rm min}(z) \equiv \left(\frac{2k T_{\rm min}}{\mu m_H G}\right)^{3/2}\left(\frac{4\pi}{3}\bar{\rho}_{\rm halo} \right)^{-1/2} \nonumber \\[10pt]
& \approx 1.6 \times 10^8\ \Msun \left(\frac{\Delta_{\rm halo}}{200}\frac{\Omega_m h^2}{0.132}\right)^{-1/2}\left(\frac{1+z}{7}\right)^{-3/2} ,
\end{align}
where the mean atomic mass is set as $\mu = 0.6$ for a fully ionized intrahalo medium. For our fiducial model assuming a universal EoR luminosity-accretion rate relation, the corresponding limiting magnitude can be fit with the linear relation,
\begin{equation}
M_{\rm UV,SF}(z) = -10.0 - 0.35(z - 6) .
\label{eqn:Muv_sf}
\end{equation}
Thus, we truncate all relevant curves in our figures at the star formation limit.

Figure \ref{fig:sfe} shows the inferred star formation efficiencies from abundance matching $\Luv$ with $M$ and with $\dot{M}$. The uncertainties in the efficiencies correspond to the those in the luminosity-mass and luminosity-accretion-rate relations shown in Figure \ref{fig:Muv}. The efficiency is not monotonic with mass nor accretion rate at any given redshift, but has a maximum value at a characteristic peak scale near where the galaxy luminosity function transitions from a power law to an exponential decline. This peak occurs at a characteristic mass $\sim 2 \times 10^{11}\ \Msun$ and a characteristic accretion rate $\sim 6 \times 10^2\ \Mdotunit$ at $z \approx 6$.

The dependence of the star formation efficiency on mass and accretion rate has a physical explanation. The reduced efficiency at higher $M$ and $\dot{M}$ is consistent with relatively inefficient atomic cooling and cold gas accretion in larger halos. The atomic cooling rate is relatively low at $T \gtrsim 10^6$ K \citep[e.g.][]{1993ApJS...88..253S} and this coincides with the observed peak mass and accretion rate. The reduced efficiency at lower $M$ and $\dot{M}$ is consistent with feedback effects from photoheating and supernova. Smaller halos are more severely affected by feedback because of the lower binding energy which scales as $M^{5/3}$. It is possible that the assumed Schechter form for the galaxy luminosity function, in particular the bright-end exponential decline and the extrapolated faint-end power law, may not be accurate and therefore would bias the inferred star formation efficiency. However, there is support for the nonmonotonic efficiency, which has been observed at lower redshifts \citep[e.g.][]{2010MNRAS.404.1111G, 2012ApJ...744..159L, 2013ApJ...770...57B}.

We find that the star formation efficiency $\varepsilon_{\rm \dot{M}}$ inferred from abundance matching $\Luv$ and $\dot{M}$ is more consistent with having no redshift evolution than the efficiency $\varepsilon_M$ inferred from abundance matching $\Luv$ and $M$. \citet{2013ApJ...762L..31B} have also found that there is lack of evolution in the star formation efficiency $\varepsilon_{\rm \dot{M}}$ for the redshift range $0 \lesssim z \lesssim 8$. \citet{2015arXiv150400005F} have shown that the stellar baryon fraction increases at higher redshifts over the range $4 \lesssim z \lesssim 8$, but this is consistent with both of our abundance matching results and does not obviously favor one particular scenario. Our results suggest that the star formation rate evolves more like $(1+z)^{2.5}$ rather than the commonly-assumed $(1+z)^{1.5}$ scaling from the 3-D analog of the Schmidt-Kennicutt relation. More precise measurements of the galaxy luminosity function and a more rigorous statistical analysis are required to strengthen this argument. 

We note that the comparison of the exponential tails in the galaxy and halo abundances is very sensitive to the value of $\Mstar$ in the Schechter luminosity function. In Figure \ref{fig:sfe}, the star formation efficiency curves are shifted to higher $M$ and $\dot{M}$ because having $\Mstar = -20.92$ at $z \approx 10$ is rather bright considering that the best-fit characteristic magnitude changed from -20.94 to -20.63 for $z \approx 6$ to 8. Recall that the extrapolation of the observed galaxy luminosity functions from a different redshift range $5 \lesssim z \lesssim 8$ yields $\Mstar \approx -20.22$ \citep{2015arXiv150601035B}. With this fainter characteristic magnitude, we expect the peaks in the star formation efficiency $\varepsilon_{\rm \dot{M}}$ in Figure \ref{fig:sfe} (right) to be even more aligned.

The efficiency at $z \approx 6$ can be fit with a triple power law,
\begin{align}
\label{eqn:sfe}
\varepsilon_{\rm \dot{M}} = 1.1 \times 10^{-2} & \left(\frac{\dot{M}}{\Mdotunit}\right)^{0.15} \nonumber \\
\times & \left(1+\frac{\dot{M}}{3.5\, \Mdotunit}\right)^{0.25} \nonumber \\
\times & \left(1+\frac{\dot{M}}{1000\, \Mdotunit}\right)^{-0.95} .
\end{align}
Equation \ref{eqn:sfe} corresponds to the best fit for the luminosity-accretion-rate relation in Equation \ref{eqn:Luv_Mdot}. This universal EoR template can be used in semi-analytical calculations and cosmological simulations to model star formation in high-redshift galaxies during the EoR. Recently, \citet{2015arXiv150700999M} have also applied the abundance matching technique to calibrate a relation between star formation rate and halo mass. They find a double power law scaling relation for $M > 10^{10}\Msun$, which is consistent with our results when the same mass range is considered.

\section{Galaxy Luminosity Function}
\label{sec:glf}

To understand how the abundance of galaxies as a function of luminosity and redshift is connected to the abundance and growth of dark matter halos, it is informative to relate the galaxy luminosity function to the halo accretion rate function and the luminosity-accretion-rate relation. The luminosity function in magnitude form can be calculated as 
\begin{equation}
\phi(\Muv, z) = \frac{dn}{d\dot{M}}\frac{d\dot{M}}{d\Muv} ,
\label{eqn:glf}
\end{equation}
where $dn/d\dot{M}$ is obtained using Equation \ref{eqn:dndmdot}. Since the luminosity-accretion-rate relation and the star formation efficiency are consistent with no evolution for the redshift range $6 \lesssim z \lesssim 10$, it is intriguing to predict the galaxy luminosity function using the universal EoR template for $\Muv(\dot{M})$ given by Equation \ref{eqn:Muv_Mdot}. If more precise measurements of the luminosity function from upcoming observations clearly deviate from this fiducial model, then it would point to exciting, additional astrophysics in star and galaxy formation.

\begin{figure}[t]
\center
\includegraphics[width=0.5\textwidth]{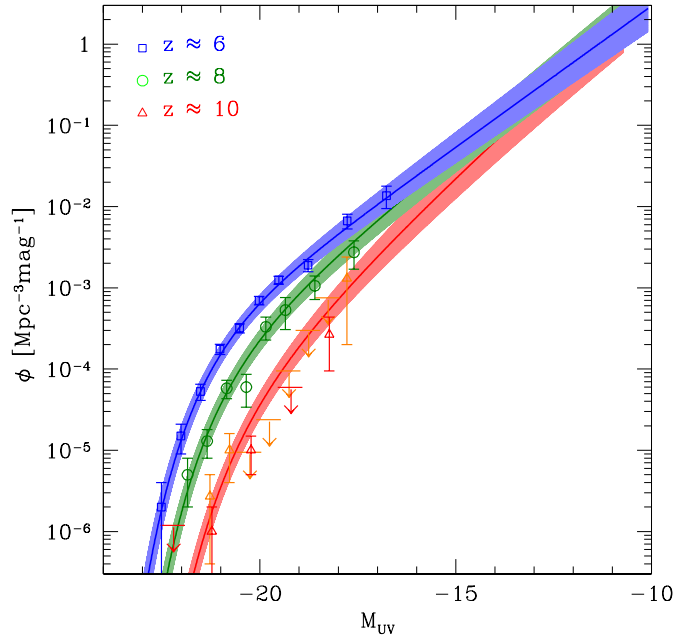}
\caption{The galaxy luminosity function at $z \approx 6$ (blue), 8 (green), and 10 (red) for our fiducial model assuming a universal EoR luminosity-accretion-rate relation. The binned observational measurements at $z \approx 6$ ( blue squares), 8 (green circles), and 10 (red triangles) from \citet{2015ApJ...803...34B} and at $z \approx 10$ (yellow triangles) from \citet{2014ApJ...786..108O} are shown for comparison. The luminosity functions match at $z \approx 6$ by construction and are in very good agreement at $z \approx 8$. At $z \approx 10$, our fiducial model is still consistent with the highly uncertain observations.}
\label{fig:glf}
\end{figure}

Figure \ref{fig:glf} shows the predicted galaxy luminosity function at $z \approx 6$, 8, and 10 for our fiducial model. The shaded prediction curves reflect the $1\sigma$ (68\%) uncertainty in the luminosity-accretion rate relation. The binned observational measurements at $6 \lesssim z \lesssim 10$ from \citet{2015ApJ...803...34B} and at $z \approx 10$ from \citet{2014ApJ...786..108O} are shown for comparison. Our results match at $z \approx 6$ by construction and are in very good agreement at $z \approx 8$. At $z \approx 10$, our model generally has larger amplitude than the few observational data points. This explains the apparent redshift dependence in the luminosity-accretion rate relation (Figure \ref{fig:Muv}) and the star formation efficiency (Figure \ref{fig:sfe}). However, our model is still consistent with the highly uncertain observations. Thus, the luminosity-accretion rate relation (Equation \ref{eqn:Muv_Mdot}) and the star formation efficiency (Equation \ref{eqn:sfe}) are highly consistent with no evolution and can be used as universal EoR templates for the EoR. \citet{2015arXiv150700999M, 2015arXiv150801204M} have also predicted similar evolution for the galaxy luminosity function by relating observed star formation rate to halo mass.

\begin{deluxetable}{rccc}[t]
\tablewidth{\hsize}
\tabletypesize{\footnotesize}
\tablecaption{\label{tab:glf} Galaxy luminosity function parameters}
\tablecolumns{4}
\tablehead{$z$ & $\phi_\star$ & $M_\star$ & $\alpha$}
\startdata
  6 & $4.9 \times 10^{-4}$ & -20.92 & -1.88 \\
  7 & $3.6 \times 10^{-4}$ & -20.76 & -1.94 \\
  8 & $2.4 \times 10^{-4}$ & -20.61 & -2.01 \\
  9 & $1.5 \times 10^{-4}$ & -20.46 & -2.08  \\
10 & $8.9 \times 10^{-5}$ & -20.32 & -2.15 \\
11 & $5.0 \times 10^{-5}$ & -20.18 & -2.22 \\
12 & $2.7 \times 10^{-5}$ & -20.04 & -2.30 \\
13 & $1.4 \times 10^{-5}$ & -19.91 & -2.38 \\
14 & $7.0 \times 10^{-6}$ & -19.78 & -2.45 \\
15 & $3.4 \times 10^{-6}$ & -19.64 & -2.53 
\enddata
\tablecomments{The best-fit Schechter parameters for our fiducial model. The normalization $\phi_\star$ has units of Mpc$^{-3}$.}
\end{deluxetable}

Table \ref{tab:glf} shows the best-fit Schechter parameters for our fiducial model for the redshift range $6 \leq z \leq 15$. At higher redshifts, the normalization $\phi_\star$ decreases dramatically, the characteristic magnitude $M_\star$ is more positive (fainter), and the faint-end slope $\alpha$ is more negative (steeper). While there are degeneracies between the best-fit parameters, all three must vary with redshift to have a good fit. Note that the differences: $\Delta\log\phi(z) = \log\phi(z) - \log\phi(6)$, $\Delta\Muv(z) = \Muv(z) - \Muv(6)$, and $\Delta\alpha(z) = \alpha(z) - \alpha(6)$ are more general and can be applied to estimate the evolution of the galaxy luminosity function when a different calibration at $z \approx 6$ is used. For better accuracy, we suggest repeating the abundance matching procedure using the new galaxy luminosity function and the halo accretion rate function.

We can explain the redshift dependence of the Schechter parameters by considering what happens to the halo accretion rate function (Figure \ref{fig:dndlgmdot}) with increasing redshift. The overall amplitude of the halo accretion rate function decreases and the normalization $\phi_\star$ also decreases accordingly. The characteristic accretion rate, at which the halo abundance undergoes exponential decline, shifts to lower values. Correspondingly, the characteristic luminosity $L_\star$ decreases and the characteristic magnitude $M_\star$ becomes more positive (fainter). Halos at a given accretion rate are rarer and the slope of the halo accretion rate function becomes steeper. Consequently, the faint-end slope $\alpha$ becomes more negative.

If more precise measurements of the galaxy luminosity function find different amplitudes at higher redshifts compared to our fiducial model, then it could be due to a number of reasons. It may simply be that the galaxy luminosity function at $z \approx 6$ used to calibrate our model may be overestimated or underestimated given the current measurement errors. Or perhaps it is due to additional astrophysics in star and galaxy formation. Lower metallicity and less metal cooling could lead to a larger Jeans mass for molecular clouds and therefore result in a more top-heavy IMF. Less photoheating and less Jeans smoothing of the IGM at an earlier stage in reionization could result in more efficient gas accretion and star formation. This would affect the formation and abundance of dwarf galaxies and lead to different values for the faint-end slope. More accurate measurements are required to understand the dependence of the star formation efficiency on mass, accretion rate, and redshift and to shed light on the galaxy formation process.

\section{Forecast for JWST}
\label{sec:jwst}

\begin{deluxetable*}{rccccrr}[th]
\tablewidth{\hsize}
\tabletypesize{\footnotesize}
\tablecaption{\label{tab:jwst} Forecast of Galaxy Counts for JWST}
\tablecolumns{7}
\tablehead{$z$ & $M_{31}$ & $M_{32}$ & $n(<M_{31})$ & $n(<M_{32})$ & $dN(<M_{31})/dz$ & $dN(<M_{32})/dz$}
\startdata
  6 & -17.9 & -16.9 & $5.1 \times 10^{-3}$ & $1.3 \times 10^{-2}$ & 50000 & 130000\\
  7 & -18.3 & -17.3 & $2.4 \times 10^{-3}$ & $6.8 \times 10^{-3}$ & 21000 & 60000 \\
  8 & -18.6 & -17.6 & $9.7 \times 10^{-4}$ & $3.2 \times 10^{-3}$ & 7800 & 26000 \\
  9 & -18.9 & -17.9 & $3.5 \times 10^{-4}$ & $1.4 \times 10^{-3}$ & 2600 & 9900 \\
10 & -19.2 & -18.2 & $1.2 \times 10^{-4}$ & $5.3 \times 10^{-4}$ & 760 & 3500 \\
11 & -19.4 & -18.4 & $3.4 \times 10^{-5}$ & $1.9 \times 10^{-4}$ & 210 & 1200 \\
12 & -19.6 & -18.6 & $8.7 \times 10^{-6}$ & $6.1 \times 10^{-4}$ & 49 & 340 \\
13 & -19.8 & -18.8 & $2.1 \times 10^{-6}$ & $1.8 \times 10^{-5}$ & 11 & 94 \\
14 & -20.0 & -19.0 & $4.2 \times 10^{-7}$ & $5.0 \times 10^{-6}$ & 2 & 24 \\  
15 & -20.2 & -19.2 & $7.8 \times 10^{-8}$ & $1.2 \times 10^{-6}$ & 0.4 & 6
\enddata
\tablecomments{The absolute magnitude $M_{\rm x}$ corresponds to an apparent magnitude limit $m_{\rm AB}=x$. Cumulative comoving number density $n(<\Mab)$ has units of Mpc$^{-3}$ and cumulative galaxy count per angular area per unit redshift $dN(<\Mab)/dz$ has units of deg$^{-2}$.}
\end{deluxetable*}

\begin{figure}[t]
\center
\includegraphics[width=0.5\textwidth]{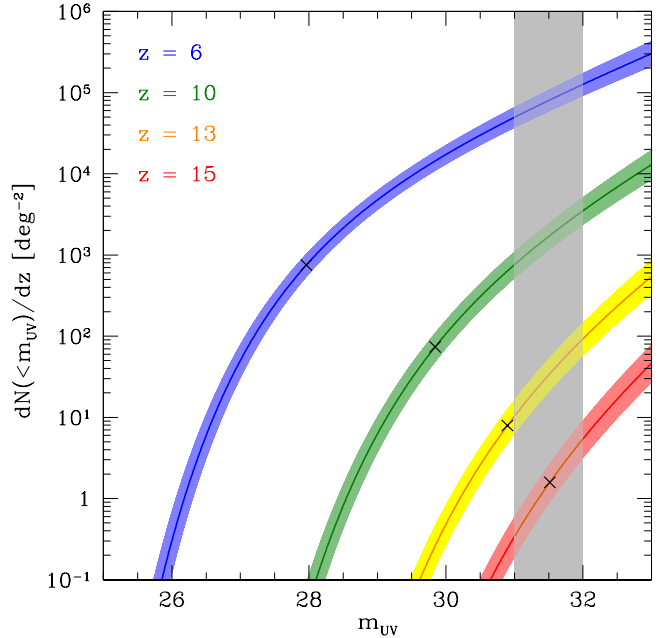}
\caption{The cumulative number of galaxies brighter than $m_{\rm UV}$ per square degree per unit redshift at $z=6$ (blue), 10 (green), 13 (yellow), and 15 (red) for our fiducial model. JWST has the sensitivity to observe unlensed galaxies at least down to $\Mstar$ (black x) at $z \lesssim 13$ (15) with apparent magnitude limit (grey) $m_{\rm AB} \approx 31$ (32).}
\label{fig:jwst}
\end{figure}

Future observations with the James Webb Space Telescope (JWST) will provide an exciting and important window to the EoR. It will allow more precise measurements of the galaxy luminosity function, particularly at lower luminosities and higher redshifts. According to \citet[][and through personal communication]{2014AAS...22324662W}, JWST has the sensitivity to reach AB apparent magnitude $m_{\rm AB} \approx 31$ with deep observations and even $m_{\rm AB} \approx 32$ with ultra deep observations. They have suggested the strategy of observing a large number of deep fields and a much larger number of medium-deep surveys on gravitational lensing foreground targets.

Figure \ref{fig:jwst} and Table \ref{tab:jwst} show the forecast based on our fiducial model for the evolution of the galaxy luminosity function. From $z=6$ to $z=15$, the limiting absolute magnitude $\Mab$ gets brighter by $\approx 2.3$ magnitudes as the luminosity distance increases by a factor of $\approx 2.8$. The comoving number density of observable galaxies $n(<\Mab)$ drops by about $4 - 5$ orders of magnitude. The number of observable galaxies per square degree per unit redshift $dN(<\Mab)/dz$ changes similarly since the differential comoving volume $dV/dz$ only changes by a factor of $\approx 2.2$. \citet{2015arXiv150700999M, 2015arXiv150801204M} have also made similar forecasts for JWST using their predictions for the evolution of the galaxy luminosity function.

JWST has the sensitivity to observe $\gtrsim 11$ (24) unlensed galaxies per square degree per unit redshift at least down to $\Mstar$ at $z \lesssim 13$ (14) with deep (ultra deep) observations. It will be able to probe some portion of the faint end at lower redshifts, but it is still about 8 (7) magnitudes away from the expected star formation limit. At $z \gtrsim 13$ (14), JWST will mainly probe the exponential tail of the luminosity function, where the rarity of galaxies will make them difficult to find. With apparent magnitude limit $m_{\rm AB} \approx 32$, it can actually reach $\Mstar$ at $z \approx 15$. Note that our forecasted counts will increase or decrease depending on how the actual galaxy luminosity function at $z \approx 6$ compares to the one used to calibrate our fiducial model.

Our fiducial model for the evolution of the galaxy luminosity function and the corresponding forecast for JWST are for our assumed cosmological parameters. If a different cosmological model is considered, then the results at $z \approx 6$ would remain the same since this is where the calibration is done, but the higher redshift results would change. For example, consider what happens if $\Omega_{\rm m}$ or $\sigma_8$ is increased. The ratio of the halo accretion rate function at redshift $z > 6$ relative to that at the pivot $z = 6$ will increase. As a result, the amplitude of the galaxy luminosity function and forecast for JWST at higher redshifts will also increase compared to our fiducial predictions.

\section{Discussion}
\label{sec:discussion}

The reionization history is determined by the evolving abundance of escaped ionization photons, which depends on the UV luminosity function, the spectral energy distribution (SED), and the radiation escape fraction of high-redshift galaxies. Only a small fraction of the ionizing photon budget comes from currently observable galaxies with $\Muv \lesssim -17$ and at $z \lesssim 10$. In order to calculate the total budget, we need to extrapolate the luminosity function to fainter magnitudes and higher redshifts. With our fiducial model, we can more confidently integrate the luminosity function because of the well-behaved and physically-motivated evolution in the Schechter parameters.

\begin{figure}[t]
\center
\includegraphics[width=0.5\textwidth]{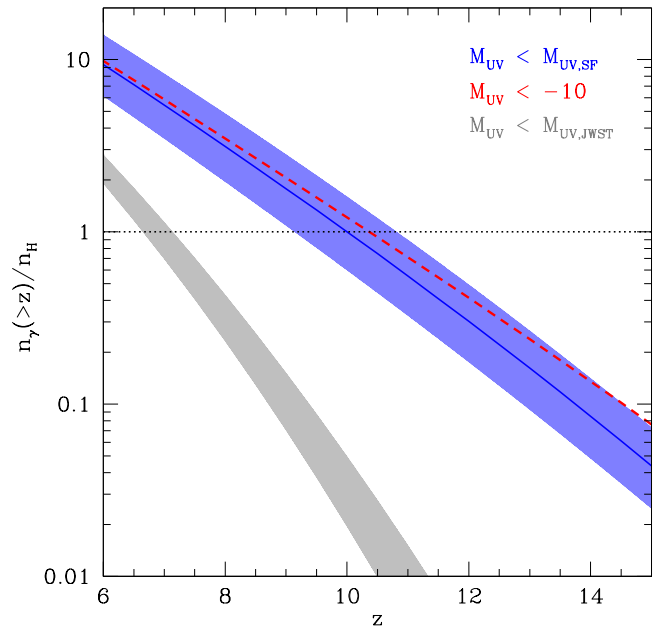}
\caption{The cumulative ionizing photon count per hydrogen atom from our fiducial model. The galaxy luminosity function is integrated down to the star formation limit (blue) and to the commonly-assumed limit $\Muv = -10$ (red dotted). Also shown is the contribution probed by JWST with apparent magnitude limit $m_{\rm UV} \approx 31-32$ (grey).}
\label{fig:reion}
\end{figure}

Figure \ref{fig:reion} shows our prediction for the cumulative ionizing photon number density,
\begin{equation}
n_\gamma(>z) = \int_z^{z_{\rm max}} \dot{n}_\gamma(z) \left|\frac{dt}{dz}\right| dz ,
\end{equation}
where the photon production rate density is calculated as
\begin{equation}
\dot{n}_\gamma(z) = \int_{M_{\rm bright}}^{M_{\rm faint}} \phi(\Muv,z) \dot{N}_\gamma(\Muv) d\Muv .
\end{equation}
A galaxy with Population II stars produces ionizing photons at a rate given by \citep[e.g.][]{2003MNRAS.344.1000B, 2003A&A...397..527S}
\begin{equation}
\dot{N}_\gamma \approx 10^{45.9 - 0.4\Muv}\, {\rm s}^{-1} \approx 10^{25.2}\ {\rm s}^{-1}\left(\frac{\Luv}{\rm erg\, s^{-1} Hz^{-1}}\right) .
\end{equation}
For the integration limits, we choose $z_{\rm max} = 25$ and $M_{\rm bright} = \Mstar - 5$. For $M_{\rm faint}$, we choose the star formation limit given by Equation \ref{eqn:Muv_sf} or the commonly-assumed limit $M_{\rm UV} = -10$ \citep[e.g.][]{2012ApJ...752L...5B, 2013ApJ...768...71R} for comparison. We also show the contribution probed by JWST with apparent magnitude limit $m_{\rm UV} \approx 31-32$.

Our fiducial model for the galaxy luminosity function at $6 \lesssim z \lesssim 8$ is in very good agreement with current observations. At $z \approx 10$, it generally has higher amplitude but shallower faint-end slope than the highly-uncertain, best-fit observations. Increasing (reducing) the ionizing photon production rate at higher redshifts would hasten (delay) the start of reionization and lengthen (shorten) its duration. Subsequently, this would affect integrated measurements of electron scattering on the CMB such as patchy Thomson scattering, kinetic Sunyaev-Zel'dovich temperature anisotropy, and polarization anisotropy. However, there is still no tension with current EoR constraints since we poorly understand how the radiation escape fraction varies with galaxy luminosity, halo mass, and redshift.

We impose a physically-motivated star formation limit based on the criterion that galaxies form in dark matter halos where the gas cools efficiently through atomic transitions. For our fiducial model, we find $M_{\rm UV,SF} \approx -10$ at $z = 6$, but it becomes more negative (brighter) with increasing redshift. This also reduces the ionizing photon production rate at higher redshifts compared to calculations using the commonly-assumed limit $\Muv = -10$ at all redshifts. Careful consideration is necessary when integrating the galaxy luminosity function down to the faintest limit, especially if the faint-end slope is steep. For example, if we extrapolate the fitting formula for the evolution of the Schechter parameters from \citet{2015ApJ...803...34B} and integrate down to the commonly-assumed magnitude limit, then the photon counts increase by more than an order of magnitude at high redshifts.

The cumulative photon count per hydrogen atom reaches unity at $z \approx 10$ and therefore, reionization is completed below this redshift since we have yet to account for the radiation escape fraction. Assuming reionization ended at $z > 6$, the luminosity-weighted radiation escape fraction is constrained by $\langle \fesc \rangle > 0.1$. The lower limit is actually higher than this because additional photons are required to balance recombinations in the clumpy IGM. Since we observe radiation escape fractions $\fesc \sim 0.01$ at $z \sim 3$ \citep[e.g][]{2006ApJ...651..688S}, this suggests that the escape fraction should vary with redshift. In addition, the average number of recombinations per hydrogen atom is $< 10$ because the escape fraction cannot exceed unity by definition. While these are only basic constraints, they already help to narrow down the parameter space of reionization. In an upcoming paper for the SCORCH project, we will make more robust constraints on the radiation escape fraction and hydrogen clumping factor.

In large-scale reionization simulations with box sizes $\gtrsim 50\ \Mpch$, radiation sources are often identified with dark matter halos from N-body simulations, with properties modeled using simple scaling relations. The source luminosity $L$ is assumed to be proportional to the halo mass $M$ through the light-to-mass ratio, which can be constant \citep[e.g.][]{2006MNRAS.369.1625Iliev}, have mass dependence \citep{2007MNRAS.377.1043McQuinn}, or have both mass and redshift dependence \citep[e.g.][]{2007ApJ...671....1T}. However, in all three cases the luminosity is deterministic, without any scatter to account for the episodic nature of starbursting galaxies. In an upcoming paper for the SCORCH project, we will use a physically-motivated and observationally-constrained approach for modeling galaxy formation in large-scale reionization simulations. Using the luminosity-accretion rate relation, we can study the effects of episodic star formation on the distribution and morphology of HII regions.

\section{Conclusions}
\label{sec:conclusions}

SCORCH is a new project to study the EoR and provide useful theoretical tools and predictions to facilitate more accurate and efficient comparison between observations and theory. In this first paper, we probe the connection between observed high-redshift galaxies and simulated dark matter halos in order to better understand the distribution and evolution of the primary source of ionizing radiation. A series of 22 high-resolution N-body simulations shown in Table \ref{tab:nbody} is used to quantify the abundance of dark matter halos as a function of mass $M$, accretion rate $\dot{M}$, and redshift $z$. The abundance matching technique is used to connect the distribution of observed high-redshift galaxies to the distribution of simulated dark matter halos. The major results are:

\begin{enumerate}
\item The halo mass function $dn/dM$ can be calculated using a self-similar barrier-crossing distribution function $f(\sigma)$ given by Equation \ref{eqn:fsigma}. The new fit is $\approx 20\%$ more accurate at the high-mass end where bright galaxies are expected to reside.
 
\item The distribution of mass accretion rate at any given mass is positively skewed. The average accretion rate $\langle\dot{M}\rangle$, variance $\mu_2$, and skewness $\mu_3$ as functions of mass and redshift are quantified by Equations \ref{eqn:mdot_avg}, \ref{eqn:mdot_var}, and \ref{eqn:mdot_ske}.

\item The halo accretion rate function $dn/d\dot{M}$ is related to the halo mass function through a mediating mass relation, and can be calculated using Equations \ref{eqn:dndmdot} and \ref{eqn:mass_mdot}. Both halo abundance functions are similar in shape, with each having a low-end power law and a high-end exponential decline.

\item The luminosity-accretion relation $\Luv(\dot{M},z)$ is more consistent with no redshift evolution than the luminosity-mass relation $\Luv(M,z)$ for the redshift range $6 \lesssim z \lesssim 10$. The former is quantified with a universal EoR template given by Equations \ref{eqn:Luv_Mdot} and \ref{eqn:Muv_Mdot}.

\item The star formation rate $\dot{M}_{\rm s}$ evolves more like $(1+z)^{2.5}$ rather than the commonly-assumed $(1+z)^{1.5}$ scaling. More precise measurements of the galaxy luminosity function and a more rigorous statistical analysis are required to strengthen this argument. 

\item The star formation efficiency $\varepsilon$ is not monotonic with mass nor accretion rate, but reaches a maximum value at a characteristic mass $\sim 2 \times 10^{11}\ \Msun$ and a characteristic accretion rate $\sim 6 \times 10^2\ \Mdotunit$ at $z \approx 6$. A corresponding template for the efficiency as a function of accretion rate is given by Equation \ref{eqn:sfe}.

\item The faintest magnitude corresponding to the star formation limit is $M_{\rm UV,SF} \approx -10$ at $z = 6$, but it becomes more negative (brighter) with increasing redshift. The redshift dependence is given by Equation \ref{eqn:Muv_sf}.

\item The fiducial model for the galaxy luminosity function is constructed from the halo accretion rate function and the luminosity-accretion-rate relation using Equation \ref{eqn:glf}. The evolution of the Schechter parameters are shown in Table \ref{tab:glf}: $\phi_\star$ decreases, $\Mstar$ is more positive (fainter), and $\alpha$ is more negative (steeper) at higher redshifts.

\item Forecast galaxy counts for JWST are shown in Table \ref{tab:jwst}. JWST has the sensitivity to observe $\gtrsim 11$ (24) unlensed galaxies per square degree per unit redshift at least down to $\Mstar$ at $z \lesssim 13$ (14) with deep (ultra deep) observations. It will also be able to probe some portion of the faint end at lower redshifts.
\end{enumerate}

The numerical fits for the abundance matching results: luminosity-accretion-rate relation, star formation efficiency, Schechter luminosity function parameters, and JWST forecasts are based on the current galaxy luminosity function at $z \approx 6$. When updated observations are available, we suggest repeating the abundance matching procedure using the new galaxy luminosity function and the halo accretion rate function.

\acknowledgments

We thank Nick Battaglia, Rychard Bouwens, Camila Correa, Jeremy Tinker, and Rogier Windhorst for helpful discussions. 
We thank Rick Costa, Roberto Gomez, and Bill Wichser for help with computing. HT acknowledges support from NASA grant ATP-NNX14AB57G and NSF grant AST-1312991. RC acknowledges support from NASA grant NNX12AF91G. Simulations were run at the NASA Advanced Supercomputing (NAS) Division, Pittsburgh Supercomputing Center (PSC), and the Princeton Institute for Computational Science and Engineering (PICSciE).

\bibliographystyle{apj}
\bibliography{scorch}

\end{document}